\documentclass[aps,prd,superscriptaddress,11pt,showpacs,notitlepage]{revtex4}
\pdfoutput=1
	\usepackage{graphicx}
	\usepackage[colorlinks=true, a4paper=true, pdfstartview=FitV,
linkcolor=blue, citecolor=blue, urlcolor=blue]{hyperref}

	\usepackage{amsmath,amssymb,mathrsfs}
\usepackage{epsf}
\usepackage{epsfig}

\usepackage{amsmath}

\usepackage{amssymb}

\numberwithin{equation}{section}

\newcommand{\be}{\begin{equation}}
\newcommand{\ee}{\end{equation}}
\newcommand{\bea}{\begin{eqnarray}}
\newcommand{\eea}{\end{eqnarray}}

\newcommand{\bb}{\bibitem}
 
\newcommand{\eqn}{\begin{eqnarray}}
\newcommand{\eqnx}{\end{eqnarray}}
\newcommand{\bA}{{\bf A}}
\newcommand{\bB}{{\bf B}}
\newcommand{\bC}{{\bf C}}
\newcommand{\ba}{{\bf a}}
\newcommand{\bd}{{\bf \delta}}

\begin{document}
\title{Exactly solvable gravitating perfect fluid solitons in $(2+1)$ dimensions}

\author{C. Adam}
\affiliation{Departamento de F\'isica de Part\'iculas, Universidad de Santiago de Compostela and Instituto Galego de F\'isica de Altas Enerxias (IGFAE) E-15782 Santiago de Compostela, Spain}
\author{T. Romanczukiewicz}
\affiliation{Institute of Physics,  Jagiellonian University,
Lojasiewicza 11, Krak\'{o}w, Poland}
\author{M. Wachla}
\affiliation{Institute of Nuclear Physics,  PAN ul. Radzikowskiego 152, 31-342 Krak\'{o}w, Poland}
\author{A. Wereszczynski}
\affiliation{Institute of Physics,  Jagiellonian University,
Lojasiewicza 11, Krak\'{o}w, Poland}

\begin{abstract}
The Bogomolnyi-Prasad-Sommerfield (BPS) baby Skyrme model coupled to gravity is considered. We show that in an asymptotically flat space-time
the model still possesses the BPS property, i.e., admits a BPS reduction to first order Bogomolnyi equations, which guarantees that the corresponding proper energy is a linear function of the topological charge. We also find the mass-radius relation as well as the maximal mass and radius. All these results are obtained in an analytical manner, which implies the complete solvability of this selfgravitating matter system. 

If a cosmological constant is added, then the BPS property is lost.  In de Sitter (\textit{dS}) space-time both extremal and non-extremal solutions are found, where the former correspond to finite positive pressure solutions of the flat space-time model. For the asymptotic anti-de Sitter (\textit{AdS}) case, extremal solutions do not exist as there are no negative pressure BPS baby Skyrmions in flat space-time.
Non-extremal solutions with \textit{AdS} asymptotics do exist and may be constructed numerically.  The impact of the negative cosmological constant on the mass-radius relation is studied. We also found two potentials for which exact multi-soliton solutions in the external \textit{AdS} space can be obtained. Finally, we elaborate on the implications of these findings for certain three-dimensional  models of holographic QCD. 
\end{abstract}

\maketitle 
%%%%%%%%%%%%%%%%%%%%%%%%%%%%%%%%%%%%%%%%%
\section{Introduction}
%%%%%%%%%%%%%%%%%%%%%%%%%%%%%%%%%%%%%%%%%
Solitons are localised finite-energy solutions of certain nonlinear field theories with a broad range of applications in high-energy physics, condensed matter physics, or even in fields like optical fibers. In several respects, solitons behave like particles. If they show up as solutions in an effective field theory (EFT) at low energies, then, frequently they may be interpreted as effective particles or quasiparticles in this effective low-energy description. One very interesting and, at the same time, quite nontrivial generalisation of solitonic field theories is their coupling to gravity, i.e., their embedding into general relativity. Depending on the context, the resulting self-gravitating soliton solutions may be interpreted as fully relativistic self-gravitating (quasi$\mbox{-}$) particles, as stars (boson stars, Skyrme stars, neutron stars, etc.) or as hairy black holes. For certain choices of the solitonic matter field theory, the gravitating soliton solutions may be interpreted as the weakly interacting gravity dual of a certain matter configuration in a strongly interacting dual QFT according to the holographic principle, shedding some light on nonperturbative properties of strongly interacting theories like QCD. Soliton models are complicated nonlinear field theories and solutions can, in general, only be found by numerical methods. The addition of gravity obviously does not simplify matters, so exact, analytical results for self-gravitating solitons may be found only in rare occasions. It is one of the main purposes of this publication to present one such rare occasion.   

One assumption which may simplify a field theory is to consider it in lower space-time dimensions.
Field theories supporting solitons in $(2+1)$ dimensions, for example, may serve as toy models for their $(3+1)$ dimensional counterparts where many ideas can be tested, leading to a deeper understanding of the higher dimensional theory.

In particular, for the Skyrme model \cite{skyrme}, which is a viable candidate for an effective theory of the strong interaction at low energy, a lower dimensional version called the baby Skyrme model \cite{baby} exists and has been widely investigated. It is given by the following Lagrange density
\be
\mathcal{L}=\frac{c_2}{2} (\partial_\mu \vec{\phi})^2 - \frac{c_4}{4} \left( \vec{\phi}_\nu \times \vec{\phi}_\rho \right)^2 -c_0 \mathcal{U} (\vec{\phi})
\ee 
where $c_2, c_4, c_0$ are positive coupling constants and $\vec{\phi}=(\phi^1, \phi^2, \phi^3)$ is an iso-vector field with unit length, i.e., $\vec{\phi} \in \mathbb{S}^2$. The potential $\mathcal{U}$ is assumed to have a discrete set of vacua, therefore static configurations can be classified by the degree of the map $\mbox{deg } [\vec{\phi}] \in \pi_2(\mathbb{S}^2)$, which is a lower dimensional version of the baryon charge assigned to Skyrmions. The model consists of three terms: {\it i)} a term quadratic in derivatives, the so-called $O(3)$ $\sigma$-model part, {\it ii)} a term quartic in derivatives, the so-called baby Skyrme part and {\it iii)} a contribution with no derivatives, that is, the potential $\mathcal{U}$. It is worth to notice that two of them, the quartic term and the potential, are obligatory from the point of view of the Derrick theorem. The $\sigma$-model term is not obligatory for the existence of solitons. It is, however, important for the behaviour of the theory close to the vacuum, determining the interactions of widely separated solitons and standing behind the complicated structure of multi-solitons \cite{multi} (chains, shells, crystals etc.). A related observation is that the baby Skyrme model can be written as a sum of two separate BPS models: the $O(3)$ $\sigma$-model
\be
\mathcal{L}_{\sigma} = \frac{c_2}{2} (\partial_\mu \vec{\phi})^2
\ee
and the BPS baby Skyrme model \cite{babyBPS}, \cite{bBPS}
\be
\mathcal{L}_{bBPS}=- \frac{c_4}{4} \left( \vec{\phi}_\nu \times \vec{\phi}_\rho \right)^2 -c_0 \mathcal{U} (\vec{\phi}).
\ee
Both models have exact solutions, for any value of the topological charge, which solve the pertinent Bogomolny equations and, therefore, saturate the corresponding topological bounds. Hence, the full baby Skyrme model can be viewed as a certain point in parameter space $\{c_0, c_2, c_4 \}$ with two distinguished limits - the BPS models \cite{stefano}. As a consequence, properties of baby Skyrmions in a given model (given values of the coupling constants) can be interpreted as resulting from a competition between these two BPS limits. In any case, understanding these two limiting models provides a solid starting point for an analytical treatment of the full model. 

Recently, the baby Skyrme model has been used as a toy model of holographic QCD \cite{sut-ads} (see also \cite{adsbaby-1}-\cite{adsbaby-3}) in the spirit of the Sakai-Sugimoto framework \cite{ss} (for recent results on holographic/\textit{AdS} Skyrmions we refer to \cite{hol}, \cite{stefano-BPS}, \cite{adsSK}). In the simplest set-up it just means putting the baby Skyrme model in an \textit{AdS} background. Because of this relation, properties of gravitating baby Skyrmions may teach us something about phases of dense nuclear matter (in particular baryonic popcorn \cite{popcorn} or dyonic salt \cite{salt}). Specifically, it has been observed that in an \textit{AdS} background a kind of popcorn phase transitions exists \cite{adsbaby-1}.   

In general, Skyrme like solitonic matter coupled to gravity forms a highly nonlinear system. Since higher charge solitons usually possess only discrete symmetries, no dimensional reduction beyond the charge one sector is possible, and full 3D or 4D numerics is required. This is a rather complicated task even if gravity is taken into account as a given background.

\vspace*{0.2cm}

In the present work, we study gravitating BPS baby Skyrmions, both in the asymptotically flat as well as \textit{dS} and \textit{AdS} space-times, where back-reaction of the matter on gravity is fully taken into account. The investigation is not only motivated by the expectation that a more analytical treatment may be possible. One should remember that purely $O(3)$ $\sigma$-model solitons in the asymptotic \textit{AdS} space-time do not exist \cite{bizon}. The existence of baby Skyrmions in \textit{AdS} space-time is, therefore, due to the BPS  part of the full baby Skyrme model. As a consequence, the analysis of the BPS baby Skyrme model can provide us with an analytical understanding of properties of baby Skyrmions in anti-de Sitter space-time, which is of relevance for the holographic toy models mentioned above. 

\vspace*{0.2cm}

The outline of the paper is as follows. In the next section we define the BPS baby Skyrme model coupled to gravity. In section III we analyze it for an asymptotically flat metric, establishing its complete solvability and its BPS nature in the full gravitating theory. We prove that all observables can be expressed via $\mathbb{S}^2$ target space integrals whose values are determined without the knowledge of a local form of the solution. Sections IV and V are devoted to de-Sitter and anti-de Sitter space-time, respectively. In particular, in Section V we show that there are no extremal (constant pressure) solutions in the asymptotic {\it AdS} case. Two exact solutions in the {\it AdS} background are also given. This is supplemented by numerical analysis of the full gravitating model, leading to some relevant implications for the existence of multi-solitons (bound states) for negative values of the cosmological constant. The last section contains our conclusions. 
%%%%%%%%%%%%%%%%%%%%%%%%%%%%%%%%%%%%%%%%%
\section{The BPS baby Skyrme model with gravity}
%%%%%%%%%%%%%%%%%%%%%%%%%%%%%%%%%%%%%%%%%
The coupling of the BPS Skyrme model with gravity is provided by  promoting the metric in the action to a dynamical variable (for convenience, we change the notation for the coupling constants $c_4, c_0$ to a new one closely related to the $(3+1)$ dimensional BPS Skyrme action \cite{BPS}), $S_{\rm tot} = S_{\rm EH} + S_{bBPS}$, where $S_{\rm EH}$ is the Einstein-Hilbert action in 2+1 dimensions with a cosmological constant (and with the
Gibbons-Hawking-York boundary term included), and
\be
S_{bBPS}=\int d^3 x |g|^{\frac{1}{2}} \left( -\lambda^2 \pi^2 |g|^{-1} g_{\alpha \beta} \mathcal{B}^\alpha \mathcal{B}^\beta - \mu^2 \mathcal{U} \right) .
\ee
Here
\be
\mathcal{B}^\mu = \frac{1}{8\pi} \epsilon^{\mu \nu \rho} \vec{\phi} \cdot \left(  \vec{\phi}_\nu \times \vec{\phi}_\rho \right)
\ee
is the topological current, where we use the same expression as in flat space for convenience (such that $\partial_\mu \mathcal{B}^\mu \equiv 0$ continues to hold, and $\mathcal{B}^\mu$ is a contravariant rank-one tensor density rather than a contravariant vector).
Further, $g_{\alpha\beta}$ is the metric tensor (we use the mostly-minus sign convention) and $g\equiv \mbox{det} g_{\alpha\beta}$.
 Then the Einstein equations are 
\be
G_{\alpha \beta}-\Lambda g_{\alpha\beta} = \frac{\kappa^2}{2} T_{\alpha \beta}
\ee
where $G_{\mu \nu}$ is the Einstein tensor and $\Lambda$ the cosmological constant. Moreover, $\kappa^2=16\pi G$ where $G$ is the gravitational constant in $(2+1)$ dimensions. 
The energy-momentum tensor is defined in the canonical way
\be
T^{\alpha \beta} = \frac{-2}{|g|} \frac{\delta S}{\delta g_{\alpha \beta} } = 2\lambda^2 \pi^2  |g|^{-1}  \mathcal{B}^\alpha \mathcal{B}^\beta - \left( \lambda^2 \pi^4 |g|^{-1} g_{\mu \nu} \mathcal{B}^\mu \mathcal{B}^\nu - \mu^2 \mathcal{U} \right) g^{\alpha \beta} 
\ee
which for the BPS baby Skyrme model leads to the perfect fluid tensor
\be
T^{\alpha \beta} = (p+\rho) u^\alpha u^\beta - p g^{\alpha \beta}
\ee
where the energy density and pressure are
\be
\rho = \lambda^2 \pi^2  |g|^{-1} g_{\mu \nu} \mathcal{B}^\mu \mathcal{B}^\nu + \mu^2 \mathcal{U} 
\ee
\be
p = \lambda^2 \pi^2  |g|^{-1} g_{\mu \nu} \mathcal{B}^\mu \mathcal{B}^\nu - \mu^2 \mathcal{U} 
\ee
while the four velocity is
\be
u^\alpha = \frac{\mathcal{B}^\alpha}{\sqrt{g_{\mu \nu} \mathcal{B}^\mu \mathcal{B}^\nu}}.
\ee
In the present paper we investigate static, axially symmetric matter. Then, only the temporal part of the topological current has a nonzero value, while $\mathcal{B}^i\equiv 0$. This assumption is compatible with a static metric $ds^2 = g_{00}(\vec x) dt^2 +g_{ij}(\vec x)dx^idx^j$, which leads to  
\be
T^{00}=\rho g^{00}, \;\;\; T^{ij}=-p g^{ij}.
\ee
Further we assume that the energy density and pressure are functions of a radial coordinate $r$ only, which is compatible with the following standard ansatz for the metric
\be
ds^2=\bA (r) dt^2 - \bB (r) dr^2 -r^2 d\varphi^2 .
\ee
Here $r$ is the geometric (or Schwarzschild-like) radius (such that spatial circles with radius $r$ have a circumference of $2\pi r$).
The nonzero components of the Einstein tensor $G_{\mu \nu}$ now read
\be
G_{00}=\frac{1}{2r} \frac{\bA \bB'}{\bB^2}, \;\;\; G_{11}= \frac{1}{2r} \frac{\bA'}{\bA}, \;\;\; G_{22} = - \frac{r^2}{4} \left( \frac{\bA'}{ \bA} \frac{\bB'}{\bB^2} + \frac{1}{\bB} \left(  \frac{\bA'^2}{ \bA^2} -  \frac{2\bA''}{ \bA}\right) \right)
\ee
where the prime denotes differentiation with respect to the radial coordinate. 
Finally, we arrive at the following set of ordinary differential equations
\eqn
\frac{\bB'}{\bB} & = & r \bB (\kappa^2 \rho+2\Lambda) \label{LeB}\\
\frac{\bA'}{\bA} & = &  r  \bB (\kappa^2 p - 2\Lambda)  \label{LeA} \\
\frac{\bA'}{\bA} \frac{\bB'}{\bB^2} +\frac{1}{\bB} \left( \frac{\bA'^2}{\bA^2} -\frac{2\bA''}{\bA} \right) &=& -2(\kappa^2 p -2\Lambda)\label{LeM}
\eqnx
where for the sake of generality we include the cosmological constant $\Lambda$, where 
\be
\rho= \frac{\lambda^2\pi^2}{\bB r^2} \mathcal{B}^0 \mathcal{B}^0 +\mu^2 \mathcal{U}
\ee
and 
\be
p= \frac{\lambda^2\pi^2}{\bB r^2} \mathcal{B}^0 \mathcal{B}^0 -\mu^2 \mathcal{U}.
\ee
The $\mathbb{S}^2$ valued matter field can be expressed by a complex scalar via stereographic projection. As a consequence, this model possesses some similarity with a (2+1) dimensional boson star built out of a complex field \cite{sakamoto}, \cite{boson star}, \cite{boson star rev}. However, in our case solitonic solutions exist also in the purely static limit. 
%%%%%%%%%%%%%%%%%%%%%%%%%%%%%%%%%%%%%%%%%
\section{Asymptotically flat metric}
%%%%%%%%%%%%%%%%%%%%%%%%%%%%%%%%%%%%%%%%%
%%%%%%%%%%%%%%%%%%%%%%%%%%%%%%%%%%%%%%%%%
\subsection{BPS property}
%%%%%%%%%%%%%%%%%%%%%%%%%%%%%%%%%%%%%%%%%
We begin our analysis with the $\Lambda=0$ case. Then the field equations are
\eqn
\frac{\bB'}{\bB} & = & \kappa^2 r \bB \rho \label{eB}\\
\frac{\bA'}{\bA} & = & \kappa^2 r  \bB p \label{eA} \\
\frac{\bA'}{\bA} \frac{\bB'}{\bB^2} +\frac{1}{\bB} \left( \frac{\bA'^2}{\bA^2} -\frac{2\bA''}{\bA} \right) &=& -2\kappa^2 p .
\eqnx
Equivalently, using $\rho-p=2\mu^2 \mathcal{U}$, the last equation can be rewritten as 
\be
(p\bB)' =  \kappa^2 r \bB^2 \mu^2 \mathcal{U} p. \label{eh}
\ee
It is straightforward to observe that the last two equations are solved for
\be
p=0, \;\;\; \bA=1,
\ee
that is, by the requirement that the matter field obeys a local zero pressure condition. Such a condition is equivalent to the fact that a solution is of the BPS type, following from a Bogomolnyi equation. Therefore, we conclude that the coupling to  gravity does not change the BPS property of this particular matter theory. Observe that this is exactly what happens with the $O(3)$ $\sigma$-model after coupling to gravity. Indeed, in asymptotically flat space $O(3)$ $\sigma$-model solitons do exist and they are solutions to a pertinent Bogomolny equation, which is just a zero pressure equation \cite{clement} (see also \cite{O(3)GR} and recent works \cite{stern 16}, \cite{stern}). This should be contrasted with the $(3+1)$ dimensional counterpart (the BPS Skyrme model) where, after coupling to gravity, the BPS property is lost \cite{BPS-star}, \cite{bjarke}. 

One should also remark that our gravitating matter falls into the framework of $(2+1)$ dimensional gravitating fluids as presented in \cite{garcia}. At a first glance, as $p=0$ everywhere, it seems a trivial case. However, this is true only if one assumes an algebraic equation of state, e.g., of the linear or polytropic form (for other algebraic EoS see \cite{rahaman}). Here, for the BPS baby Skyrme model, the local equation of state describes a non-barotropic  fluid and, therefore, does not have an algebraic form. Specifically, zero pressure can correspond to a nontrivial energy and particle distribution. Furthermore, the fact that we get a local zero pressure solution circumvents an observation that for vanishing cosmological constant a gravitating fluid represents a cosmological solution and there is no surface of vanishing pressure \cite{garcia}. 

There are two more equations which remain to be solved
\eqn
\frac{\bB'}{\bB} & = & \kappa^2 r \bB \rho \\
\rho&=&2\mu^2 \mathcal{U} \label{bps-gr}
\eqnx
where the second one is just the zero pressure condition.  Now, we express the $\vec{\phi}$ field by a complex scalar using the standard stereographic projection
\be
\vec{\phi}=\frac{1}{1+|u|^2} \left( u+\bar{u}, -i ( u-\bar{u}),
1-|u|^2 \right)
\ee
and apply the following ansatz 
\be 
u=f(r)e^{in\varphi} .
\ee
Hence,
\be
\mathcal{B}^0 = \frac{1}{8\pi} \epsilon^{ij} \vec{\phi} \cdot (\vec{\phi}_i \times \vec{\phi}_j) = \frac{1}{2\pi i} \epsilon^{ij}\frac{u_i \bar{u}_j}{(1+|u|^2)^2} =-\frac{n}{\pi} \frac{ff_r}{(1+f^2)^2} .
\ee
If the profile function $f$ obeys the proper boundary conditions 
\be
f(r=0)=\infty, \;\;\; f(R)=0
\ee
then the configuration has topological charge $n$. Here, $R$ is the geometric radius of the soliton which can be finite (compactons \cite{compacton}) or infinite (usually infinitely extended solitons). In flat space, the importance of this geometrical radius is related to the fact that it defines a geometrical volume which is, at the same time, the proper thermodynamical volume. 
Following our parametrization 
\be
\rho = \frac{\lambda^2n^2}{\bB r^2} \frac{f^2f_r^2}{(1+f^2)^4} +\mu^2 \mathcal{U}(f), \;\;\; p = \frac{\lambda^2n^2}{\bB r^2} \frac{f^2f_r^2}{(1+f^2)^4} -\mu^2 \mathcal{U}(f).
\ee
Then, the matter equation i.e., $p=0$, reads explicitly 
\be \label{matt-eq}
 \frac{\lambda^2n^2}{\bB r^2} \frac{f^2f_r^2}{(1+f^2)^4} =\mu^2 \mathcal{U}(f),
\ee
which may be transformed into the form of the flat space BPS equation by a suitable change of the radial variable. In a first step, we introduce the proper radius $\varrho$ defined by
\be
d\varrho = \sqrt{\bB} dr \quad \Rightarrow \quad \varrho (r) = \int_0^r dr' \sqrt{\bB(r')}
\ee
which leads to the metric
\be
ds^2 = \tilde \bA (\varrho ) dt^2 - d\varrho^2 - \tilde \bC (\varrho) d\varphi^2
\ee
 and directly measures radial distances. In flat space, it is useful to introduce the "quadratic" coordinate $z=(1/2) r^2$. In our case, the correct generalisation which makes the factor $\bB$ disappear from Eq. (\ref{matt-eq}) is 
 \be \label{z}
   dz = \sqrt{\bB} rdr = rd\varrho
 \quad \Rightarrow \quad z= \int_0^r dr' r'\sqrt{\bB (r')}.
 \ee
 For a geometrical understanding, observe that in the corresponding proper volume element $ dz d\varphi =   d\varrho \, rd\varphi$, $d\varrho$ measures proper lengths in the radial direction, whereas $r d\varphi$ measures proper lengths in the angular direction.
 
In the new variable $z$, Eq. (\ref{matt-eq}) reads
\be
\lambda^2n^2\frac{f^2f_z^2}{(1+f^2)^4} =\mu^2 \mathcal{U}(f), \label{BOG-f}
\ee
that is, the usual Bogomolny equation for the BPS baby Skyrme model without gravity \cite{babyBPS}.
In the last step, we simplify this Bogomolny equation, introducing a new target space variable 
\be 
h=1-\frac{1}{1+f^2} \label{h}.
\ee
Then,
\be
 \frac{\lambda^2n^2}{4} h'^2- \mu^2 \mathcal{U}(h) =0 ,\label{BOG}
\ee
and the boundary conditions for $h$ are
\be
h(z=0)=1, \;\;\; h(z=z_0)=0, \;\;\; h_z(z=z_0)=0.
\ee
Here, $z_0= z(R) = \int_0^R r'd\varrho$, where $R$ is the geometric radius of the solution. Topologically nontrivial solutions of this equation for several classes of potentials have been studied in \cite{babyBPS}.

Finally, we rewrite the equation for the metric function $\bB$ (\ref{eB}) in the new variable $z$ as
\be
\frac{\bB'_r}{r\bB^2} = \kappa^2 \rho(r) \;\; \Rightarrow \;\;  \frac{\bB'_z}{\bB^{3/2}} =  \kappa^2 \rho (z)  \;\; \Rightarrow \;\;  \frac{d }{dz} \bB^{-1/2} =  -\frac{1}{2} \kappa^2 \rho (z).
\ee
This is easily solved with the following condition
\be
\bB(z=0)=	1
\ee
i.e., flat metric in the origin. Let us remark that in $(2+1)$ dimensions there is no local definition of mass and, therefore, no unique choice of the value of the metric at the origin. Our choice is motivated by the requirement that there is no
conical singularity of the spatial metric at the origin \cite{sakamoto}. Hence, 
\be
\bB^{-1/2}(z)= 1-\frac{1}{2} \kappa^2 \int_0^z \rho (z') dz' \label{B-z}.
\ee
As the metric function $\bB$ has to be non-singular, we get a restriction 
\be
\frac{1}{2} \kappa^2 \int_0^{z_0} \rho (z) dz =\kappa^2 \mu^2 \int_0^{z_0} \mathcal{U} (h(z)) dz   <1 \label{n-bound1}
\ee
which for fixed, given values of the coupling constants leads to a condition on the topological charge. 

%%%%%%%%%%%%%%%%%%%%%%%%%%%%%%%%%%%%%%%%%
\subsection{Linear observables}
%%%%%%%%%%%%%%%%%%%%%%%%%%%%%%%%%%%%%%%%%

There are two quantities which may characterise these gravitating solitons in a geometric way, namely,  its proper mass and volume. Both observables can be computed in the model {\it without} finding a particular matter solution. What is needed is just the Bogomolny equation. Let us start with the proper mass, which is the volume integral of the proper energy density 
\be
M=\int d^2 x |g|^{\frac{1}{2}} \rho (\vec{x}) = 2\pi \int_0^\infty dr r \sqrt{\bB} \left( \frac{\lambda^2n^2}{\bB r^2} \frac{f^2f_r^2}{(1+f^2)^4} +\mu^2 \mathcal{U}(f)\right)
\ee
where we used the fact that $\bA=1$ and the $\varphi$ independence of the energy density. Note, that it differs from the asymptotic or ADM mass. In fact, the proper mass is just the non-gravitational part of the total (ADM) mass. Now, using the variables $h$ and $z$, 
\be
M=2\pi \int_0^\infty dz \left( \frac{\lambda^2n^2}{4} h_z^2 +\mu^2 \mathcal{U}(h)\right) = 2\pi \lambda\mu |n|  \int_0^\infty dz h_z \sqrt{\mathcal{U}} 
\ee
where the Bogomolny equation (\ref{BOG}) has been applied. This can be written as a target space (solution independent) expression \cite{martin}
\be
M=2\pi \lambda\mu |n|  \int_0^{1} dh \sqrt{\mathcal{U}(f)}   = 2\pi \lambda\mu |n|  \langle \sqrt{\mathcal{U}} \rangle_{\mathbb{S}^2} \label{bps-M}
\ee
where $\langle \sqrt{\mathcal{U}} \rangle_{\mathbb{S}^2}$ is the $\mathbb{S}^2$ target space average (here ${\bf vol}_{\mathbb{S}^2}$ is the area two-form on $\mathbb{S}^2$) 
\be
\langle \sqrt{\mathcal{U}} \rangle_{\mathbb{S}^2} \equiv 
\frac{\int_{\mathbb{S}^2} {\bf vol}_{\mathbb{S}^2} \sqrt{\mathcal{U}}}{\int_{\mathbb{S}^2} {\bf vol}_{\mathbb{S}^2} } =
2 \int_0^{\infty} df \frac{f \sqrt{\mathcal{U}(f)}}{(1+f^2)^2}   =   \int_0^{1} dh \sqrt{\mathcal{U}(h)} .
\ee
Hence, our condition (\ref{n-bound1}) can be written as
\be
\frac{\kappa^2}{2} \frac{M}{2\pi } = \frac{1}{2} \kappa^2 \lambda \mu |n| \langle \sqrt{\mathcal{U}} \rangle_{\mathbb{S}^2}<1
\ee
or
\be
 |n| < \frac{1}{\kappa^2 \lambda \mu} \frac{2}{  \langle \sqrt{\mathcal{U}} \rangle_{\mathbb{S}^2}}   .\label{n-bound2}
\ee
Note that the mass is still proportional to the topological charge and, therefore, the model is a true BPS model even after the coupling with gravity. This fact differs from the $(3+1)$ dimensional counterpart, i.e., the BPS Skyrme model, where gravity destroys the BPS property. As we have seen, the maximal topological charge $n^{max}$ of the soliton is derived from the maximal mass condition. Hence,
\be \label{nmax}
n^{max}=  \frac{1}{\kappa^2 \lambda \mu} \frac{2}{  \langle \sqrt{\mathcal{U}} \rangle_{\mathbb{S}^2}}, \;\;\;\; M^{max} = \frac{4\pi}{\kappa^2} .
\ee

In a similar fashion, we can compute the geometric proper volume (more precisely, area; but we shall continue to use the generic notion of "volume")
 \be
 V=\int d^2 x \sqrt{g^{(2)}} = 2\pi \int_0^R dr r\sqrt{\bB} =2\pi  \int_0^{z_0}dz
 \ee
 ($g^{(2)} \equiv \det g_{ij}$)
 where $R$ is the compacton radius. From the BPS equation in the $z$ coordinate we find that (we take the minus sign)
 \be
 dz = -\frac{n\lambda }{2\mu} \frac{dh}{\sqrt{\mathcal{U}}} \label{dz}
 \ee
 and 
 \be
 V=2\pi \int_0^{z_0}dz= \pi \frac{\lambda }{\mu} |n| \int_0^1 \frac{dh}{ \sqrt{\mathcal{U}}} = \pi \frac{\lambda }{\mu} |n|  \left\langle \frac{1}{ \sqrt{\mathcal{U}}} \right\rangle_{\mathbb{S}^2} \label{bps-V}
 \ee
 Finally,  we get a linear relation between the proper mass and proper volume 
\be
M=2\mu^2 \frac{ \langle \mathcal{U}^{1/2} \rangle_{\mathbb{S}^2}}{ \langle \mathcal{U}^{-1/2} \rangle_{\mathbb{S}^2}} V = \frac{4\pi}{\kappa^2} \frac{V}{V^{max}},  \label{proper-MR}
\ee
where the maximal proper volume is 
\be
V^{max}=\frac{4\pi}{\kappa^2} \frac{ \langle \mathcal{U}^{-1/2} \rangle_{\mathbb{S}^2}}{ \langle \mathcal{U}^{1/2} \rangle_{\mathbb{S}^2}}.
\ee
Note that, as expected for a BPS theory, the proper mass as well as the volume are {\it linear} functions of the topological charge, with the coefficients given by target space averages of functions of the potential (\ref{bps-M}), (\ref{bps-V}). 

It should be underlined that the geometric volume is finite only if the model, i.e., used potential, leads to compact solitons \cite{martin}. This requires that close to the vacuum at $h=0$ the potential behaves as $\mathcal{U} \sim  h^a$ with $a<2$. For other types of potentials gravitating solitons can perfectly exist but they are infinitely extended objects. This requires another definition of the volume (radius). For example, one can use the mean square radius. However, the corresponding volume lacks uniqueness and a thermodynamical interpretation.  

We remark that the compacton nature of the gravitating solitons is inherited from the underlying non-gravitating theory (for examples of compact boson stars see \cite{jutta}). 
%%%%%%%%%%%%%%%%%%%%%%%%%%%%%%%%%%%%%%%%%
\subsection{Nonlinear observables and mass-radius relation}
%%%%%%%%%%%%%%%%%%%%%%%%%%%%%%%%%%%%%%%%%
The expressions (target space integrals) for proper mass and volume are identical to the non-gravity case. Hence, the obtained exact formulas follow simply from previous works. In the case with gravity, however, we will find that also the geometric radius $R$ as well as the total (asymptotic) mass can be found without the knowledge of a particular solution, i.e., expressed entirely by means of target space integrals. 

The ADM (total or asymptotic) mass $M_{ADM}$ in $(2+1)$ dimensions is defined by an asymptotic (constant) component of a metric function \cite{boson star}, \cite{betti}. For our choice of the metric the definition is
\be
\frac{1}{\bB (r)} = 1 - 2m(r) \;\;\;\mbox{and} \;\;\;  M_{ADM} =  \frac{4\pi} {\kappa^2 } \lim_{r\rightarrow \infty } m(r)
\ee
or
\be
M_{ADM}=\frac{4\pi}{2\kappa^2} \left( 1 - \lim_{r\rightarrow \infty} \frac{1}{\bB (r)} \right)
\ee
However, using eq. (\ref{B-z}) we get that 
\be
\lim_{r\rightarrow \infty}  \bB^{-1}  = \left( 1-\frac{1}{2} \kappa^2 \int_0^{z_0} \rho (z) dz \right)^2 = \left(1-\frac{\kappa^2}{4\pi} M  \right)^2 .
\ee
Thus finally, 
\be
M_{ADM}= M \left( 1-\frac{\kappa^2}{8\pi} M \right)= M \left( 1-\frac{1}{2} \frac{M}{M^{max}}  \right)
\ee
or, after using the exact expression for the proper mass, 
\be
M_{ADM}=2\pi \lambda\mu |n|  \langle \sqrt{\mathcal{U}} \rangle_{\mathbb{S}^2} \left(1- \frac{1}{4} \kappa^2 \mu \lambda |n| \langle \sqrt{\mathcal{U}} \rangle_{\mathbb{S}^2} \right). \label{ADM-mass} 
\ee
The total mass is not a linear function of the topological charge. Owing to the gravitational interaction, it receives a negative correction which is quadratic in the topological charge.
One should remember that there could emerge an instability due to the attractive gravitational interaction which means that for sufficiently large topological charge the total mass decreases with $|n|$. Hence, we impose that
\be
\frac{dM_{ADM}}{dn} > 0 \;\;\; \Rightarrow \;\;\; n<n^*\equiv \frac{2}{\kappa^2 \mu \lambda \langle \sqrt{\mathcal{U}} \rangle_{\mathbb{S}^2} } 
\;\;\; \mbox{or} \;\;\; M< M^{max} .
\ee 
Thus, the instability condition is exactly the same as the former $n^{max}$ condition (\ref{nmax}). As a consequence, the maximal total mass is 
\be
M_{ADM}^{max}=M_{ADM}(M^{max})=\frac{1}{2} M^{max} 
\ee
which is one-half of the maximal proper mass. 
Note also that both, maximal total as well as proper mass, do not depend on a particular form of the  potential. It is also worth observing that the total mass can be found as the following integral
\be
M_{ADM}= 2\pi \int_0^\infty dr r \rho(r) .
\ee
Indeed,
\bea
2\pi \int_0^\infty dr r \rho(r) &=& \\
&=& 2\pi \int_0^{z_0} dz \bB^{-1/2}(z)  \left( \frac{\lambda^2n^2}{4} h_z^2  +\mu^2 \mathcal{U}(h)\right)\\
& =& - 2\pi \mu \lambda |n| \int_0^{z_0} dz \bB^{-1/2}(z) h_z \sqrt{\mathcal{U}} \\
&=&  - 2\pi \mu \lambda |n| \int_0^{z_0} dz  h_z \sqrt{\mathcal{U}} \left( 1-\frac{1}{2} \kappa^2 \int_0^z \rho (z') dz' \right) \\
&=& 2\pi \mu \lambda |n| \int_0^1dh \sqrt{\mathcal{U}} \left( 1-\frac{\kappa^2}{2} \lambda\mu |n| \int^1_h \sqrt{\mathcal{U}} dh'\right) \\
\eea
where we used the Bogomolny equation and equation (\ref{B-z}). Thus,
\be
M_{ADM}=2\pi \lambda\mu |n|  \langle \sqrt{\mathcal{U}} \rangle_{\mathbb{S}^2} - \pi \kappa^2 \mu^2 \lambda^2 n^2 \int_0^1dh\left(  \sqrt{\mathcal{U}(h)}  \int^1_h \sqrt{\mathcal{U}(h')} dh' \right)
\ee
where $n< n^{max}$. However, the double integral can be simplified to the following form 
\be
\int_0^1dh\left(  \sqrt{\mathcal{U}(h)}  \int^1_h \sqrt{\mathcal{U}(h')} dh' \right) =\frac{1}{2} \langle \sqrt{\mathcal{U}} \rangle_{\mathbb{S}^2} ^2
\ee
and we arrive at the above definition of the total mass. 

We remark that in $(2+1)$ dimensional gravity (especially with a negative cosmological constant) another mass definition is frequently used. However, it is just ADM mass minus a constant. Therefore, it is obviously computable in the same fashion as the ADM mass. There is another interesting quantity often considered for topological solitons. Namely, the binding energy which measures, for example, how strong two charge one solitons are bound into a single charge two object. We define the excess energy (that is, minus the binding energy)
\be
\Delta M_{ADM} = M_{ADM} (n=2) -   2 M_{ADM}(n=1).   
\ee
Using formula (\ref{ADM-mass}) we find that
\be
\Delta M_{ADM} = - \kappa^2 \lambda^2 \mu^2 \pi  \langle \sqrt{\mathcal{U}} \rangle_{\mathbb{S}^2} ^2 .
\ee
The expression is always negative. Owing to the attractive gravitational interaction, the ADM mass of the charge two soliton is smaller than two charge one baby Skyrmions. The binding energy obviously vanishes in the no-backreaction limit. Similarly, one can obtain the relative excess energy $\delta_2 = \Delta M_{ADM}/(2M_{ADM}(n=1))$. Note that the value as well as  the sign of the binding energy is rather sensitive to a particular choice of the mass definition. 

\begin{figure}
\hspace*{-1.0cm}
\includegraphics[height=5cm]{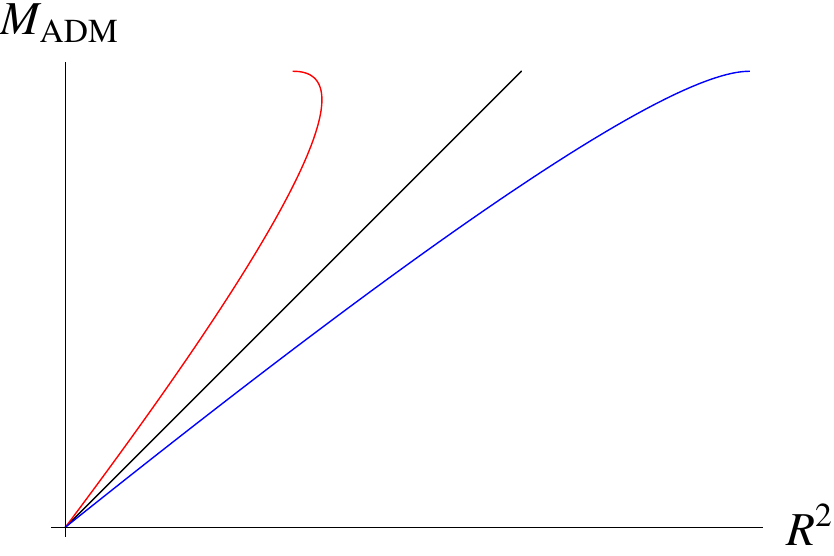}
\includegraphics[height=5.cm]{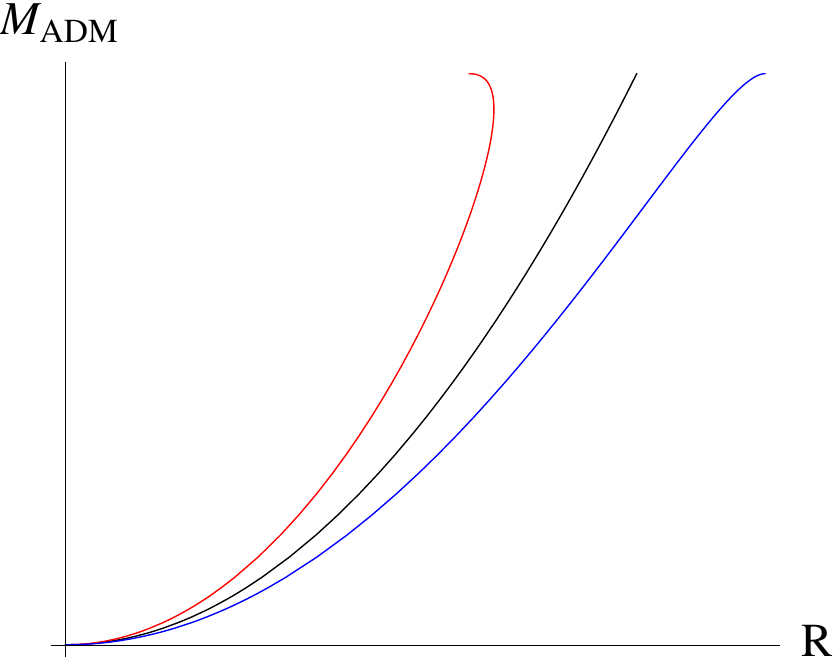}
\caption{Left: Schematic ADM mass-volume (radius squared) curves for the BPS baby Skyrme model for $\Omega<2$ (red), $\Omega=2$ (black) and $\Omega>2$ (blue). Right: corresponding $M$-$R$ curves.}
\label{MR2-O}
\end{figure}

Finally, the geometric radius can be computed as (we use (\ref{dz}) which allows to change any integral over the base space for an integral over the target space)
\be
R^2 = \frac{\lambda}{\mu} |n| \int_0^1 \frac{dh}{\sqrt{\mathcal{U}}} \left( 1-\frac{\kappa^2}{2} \lambda\mu |n| \int^1_h \sqrt{\mathcal{U}} dh'\right) \label{Rsol}
\ee
or
\be
R^2= \frac{\lambda }{\mu} |n| \left( \left\langle \frac{1}{ \sqrt{\mathcal{U}}} \right\rangle_{\mathbb{S}^2}  - \frac{1}{2} \kappa^2 \mu \lambda |n|   \int_0^1\frac{1}{ \sqrt{\mathcal{U}(h)}} \int_h^1 \sqrt{\mathcal{U}(h')} dh'  \right) .
\ee
Again, due to the gravitational interaction the radius gets a contribution which is quadratic in the topological charge. The maximal radius occurs for
\be
n^{R \; max}= \frac{1}{\kappa^2 \mu \lambda} \left\langle \frac{1}{ \sqrt{\mathcal{U}}} \right\rangle_{\mathbb{S}^2} \left(  \int_0^1\frac{1}{ \sqrt{\mathcal{U}(h)}} \int_h^1 \sqrt{\mathcal{U}(h')} dh' \right)^{-1}
\ee
and reads
\be
R^2_{max} =\frac{\lambda}{\mu}  \frac{n^{R\; max}}{2}  \left\langle \frac{1}{ \sqrt{\mathcal{U}}} \right\rangle_{\mathbb{S}^2}= \frac{1}{2 \kappa^2 \mu^2 } \left\langle \frac{1}{ \sqrt{\mathcal{U}}} \right\rangle_{\mathbb{S}^2}^2 \left(  \int_0^1\frac{1}{ \sqrt{\mathcal{U}(h)}} \int_h^1 \sqrt{\mathcal{U}(h')} dh' \right)^{-1}
\ee
It is a rather amazing feature of this gravitating matter theory that {\it all} observables can be found in closed form without solving the underlying equations (without knowledge of a local solution). Furthermore, we can use the exact formulas for the total mass and radius to get, after eliminating the topological charge $n$, a general (uniformized) mass-radius curve. In a parametric way, where $x=|n|/n^{max} \in [0,1]$, it reads
\be
\left\{
\begin{array}{l}
\frac{\kappa^2 M_{ADM}}{2\pi } = x  \left( 2 -x \right)\\
\\
\frac{\kappa^2 \mu^2 R^2}{2}= \frac{\mathcal{A}}{\langle \sqrt{\mathcal{U}} \rangle_{\mathbb{S}^2}^2 }  x \left( \frac{ \langle \mathcal{U}^{1/2}\rangle_{\mathbb{S}^2}  \left\langle \mathcal{U}^{-1/2} \right\rangle_{\mathbb{S}^2} }{\mathcal{A}} - x  \right)
\end{array}
\right.
\ee
where we introduced the constant 
\be
\mathcal{A}\equiv \int_0^1\frac{1}{ \sqrt{\mathcal{U}(h)}} \int_h^1 \sqrt{\mathcal{U}(h')} dh'
\ee
which together with $\langle \sqrt{\mathcal{U}} \rangle_{\mathbb{S}^2}$, $\langle 1/\sqrt{\mathcal{U}} \rangle_{\mathbb{S}^2}$ defines a particular shape of the mass-radius curve. Note that the qualitative shape of the mass-radius curve depends on the following ratio  
\be
\Omega \equiv  \frac{ \langle \mathcal{U}^{1/2}\rangle_{\mathbb{S}^2}  \left\langle \mathcal{U}^{-1/2} \right\rangle_{\mathbb{S}^2} }{\mathcal{A}}  \label{Omega}
\ee
It is especially clearly visible for the mass-volume (radius squared) curve. Indeed, 
if $\Omega <2$ we get a mass-volume curve which turns back at a certain $R_{max}$. For $\Omega=2$ the curve is just a straight line. Finally, for $\Omega>2$ the curve gets flattened for large radii and the $R_{max}$ point coincides with $M_{max}$, both occurring for $n=n^{max}$. This is schematically shown in Fig. \ref{MR2-O}, left panel. Note that in the plot the units depend on a particular choice of the potential and, therefore, the slope of the curve is not properly visible. In the right panel we show the corresponding $M$-$R$ curves.  

Note that in the $\Omega=2$ case the relation between the ADM mass and radius is {\it exactly} the same as the relation between the proper mass and the proper radius (\ref{proper-MR}). The only difference is that the maximal ADM mass is half of the maximal proper mass. Similarly, these curves are identical to the mass-radius relation in the no gravity limit ($\kappa=0$) which, of course, continue up to arbitrarily high masses (topological charges). 

In general, one can prove that $\Omega \geq 1$ whenever all integrals entering the definition of $\Omega$ exist, which means that the mass-radius curve is well defined for any potential with compact solitons. That is, $R >0$ for any $n \leq n_{max}$.

One can also easily verify that for the maximally stiff matter i.e., for the constant potential $\mathcal{U}(h)=U_0 \Theta (h)$ (where $\Theta$ is the Heaviside step function and $U_0$ a number) we get $\Omega=2$. Furthermore, for many typical, and physically well motivated, one vacuum potentials (for example $\mathcal{U}=U_0 h^a, a\in[0,2)$) we find $\Omega<2$. However, it is possible to construct a potential with $\Omega > 2$. For instance one may consider $\mathcal{U}=(h+1/2)^{-1} \Theta(h)$ which leads to $\Omega=4$. Nonetheless, taking into account the rather unusual form of such a potential, we can conclude that models with $\Omega<2$ seem to be more plausible from a physical point of view.  

For potentials with $\Omega <2$, the strength of the bending towards the left can be measured by the difference between $R_{max}$ and $R(n_{max})$
\be
\frac{\kappa^2\mu^2}{2} \left( R_{max}-R(n_{max}) \right)= \frac{1}{\Omega} \frac{ \langle \mathcal{U}^{-1/2}\rangle_{\mathbb{S}^2} }{ \langle \mathcal{U}^{1/2}\rangle_{\mathbb{S}^2} } \left( \frac{\Omega}{2} -1\right)^2  \label{DeltaR}
\ee
or simply by the radius at $n_{max}$
\be
\frac{\kappa^2\mu^2}{2}  R_{n_{max}} =  \frac{ \langle \mathcal{U}^{-1/2}\rangle_{\mathbb{S}^2} }{ \langle \mathcal{U}^{1/2}\rangle_{\mathbb{S}^2} } \left(1- \frac{1}{\Omega } \right)  \label{Rn_max}
\ee

At the end, we comment that any deformation of this qualitative picture may be used as a signature of a departure from the Einstein gravity. 
%%%%%%%%%%%%%%%%%%%%%%%%%%%%%%%%%%%%%%%%%
\subsection{Examples}
%%%%%%%%%%%%%%%%%%%%%%%%%%%%%%%%%%%%%%%%%
%%%%%%%%%%%%%%%%%%%%%%%%%%%%%%%%%%%%%%%%%
\subsubsection{The old baby potential}
%%%%%%%%%%%%%%%%%%%%%%%%%%%%%%%%%%%%%%%%%
As an example, we consider the so-called old baby potential 
\be
\mathcal{U}_\pi=\frac{1}{8}(1-\phi^3)=\frac{h}{4} . \label{old}
\ee
This is the most known and most used potential in physical applications of the baby Skyrme model, with multi-soliton configurations forming chains. First of all, for this potential
\be
\langle \sqrt{\mathcal{U}} \rangle_{\mathbb{S}^2}= \frac{1}{3}, \;\;\; \left\langle \frac{1}{\sqrt{\mathcal{U}}} \right\rangle_{\mathbb{S}^2}=4 \;\;\; \mbox{and} \;\;\; \mathcal{A}=1.
\ee
Thus,
\be
M=\frac{2\pi}{3} \mu \lambda |n|, \;\;\; V=4\pi \frac{\lambda}{\mu} |n| .
\ee
Now we consider the BPS equation in the new coordinate $z$ 
\be
 \lambda^2n^2 h'^2_z= \mu^2 h
\ee
which leads to the following compacton solution (remember $V=2\pi z_0$)
\be
h= \left\{
\begin{array}{ccl}
\left(1-\frac{z}{z_0} \right)^2 & & z\leq z_0=\frac{2\lambda n}{\mu} \\
0 & & z \geq z_0 .
\end{array} \right.
\ee
From this exact matter solution we can find the metric function, again in the $z$ variable, as
\be
\bB^{-1/2}(z)=1-\frac{\kappa^2 \mu^2 z_0}{12} + \frac{\kappa^2 \mu^2 z_0}{12}   \left(1-\frac{z}{z_0} \right)^3  
\ee
This gives the following condition
\be
1> \frac{\kappa^2 \mu^2 z_0}{4\cdot 3} \;\; \Rightarrow \;\; |n|<n^{max}= \frac{6}{\kappa^2 \mu \lambda }
\ee
which obviously agrees with the general formula. Furthermore, the explicit metric solutions allow us to find an exact relation between the $z$ and $r$ variables
\be
\int_0^{z} \frac{dz'}{\sqrt{\bB(z')}} = \int_0^r r'dr' = \frac{r^2}{2}
\ee
i.e.,
\be
\frac{r^2}{2} = \left(1-\frac{\kappa^2 \mu^2 z_0}{12} \right)z - \frac{\kappa^2 \mu^2 z_0^2}{4\cdot 12}  \left(1-\frac{z}{z_0} \right)^4 +\frac{\kappa^2 \mu^2 z_0^2}{4\cdot 12}
\ee
Then, the radius is 
\be
\frac{R^2}{2} = \left(1-\frac{\kappa^2 \mu^2 z_0}{16} \right)z_0 \;\; \Rightarrow \;\;
R^2 = 4\frac{\lambda }{\mu} |n| \left(1-\frac{1}{8} \kappa^2 \lambda \mu |n| \right) = 4\frac{\lambda }{\mu} |n| \left(1-\frac{3}{4}\frac{|n|}{n^{max}} \right)  \label{R flat}
\ee
\begin{figure}
\hspace*{-1.0cm}
\includegraphics[height=5.cm]{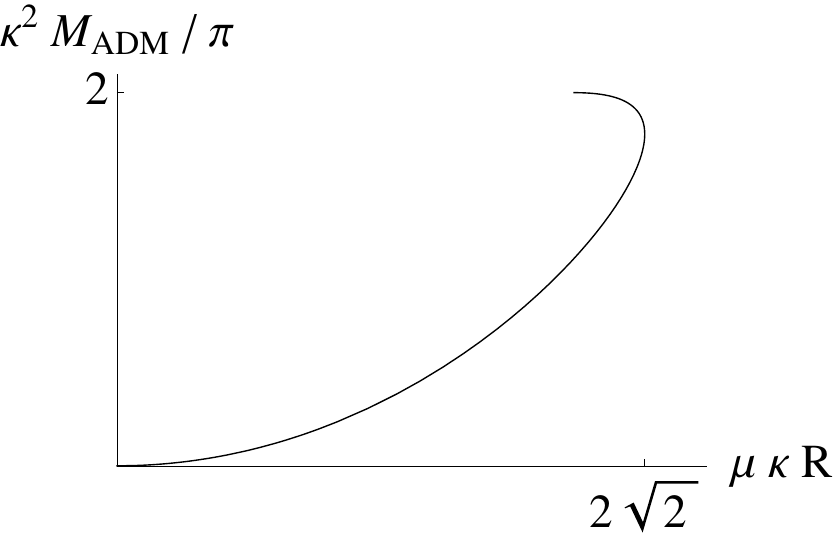}
\caption{Asymptotic mass-radius curve for the BPS baby Skyrmions with the old baby potential $\mathcal{U}_\pi$ in the asymptotically flat space-time.}
\label{MR}
\end{figure}
This agrees with the general formula (\ref{Rsol}), as it should. Similarly, we can compute the total mass of the gravitating axially symmetric soliton 
\be
M_{ADM}=\frac{2\pi}{3} \lambda \mu |n| \left( 1-\frac{1}{12} \kappa^2 \mu \lambda |n|\right) =\frac{2\pi}{3} \lambda \mu |n| \left(1-\frac{1}{2}\frac{|n|}{n^{max}} \right) . \label{m-tot flat}
\ee
Now we derive an exact relation between the radius and the total mass of the gravitating baby Skyrmion given by the parametric set of equations (\ref{R flat}, \ref{m-tot flat}). As previously mentioned it corresponds to the $\Omega <2$ type. Specifically, $\Omega=4/3$. The maximal total mass obviously occurs for $|n|=n^{max}$ and is one-half of the maximal proper mass. 
On the other hand, the maximal radius is found for $|n|=\frac{2}{3}n^{max}$ 
\be
R^{max}=\frac{2\sqrt{2}}{\kappa \mu} .
\ee
As we see in Fig. \ref{MR}, the resulting curve is very similar to the one found in the $(3+1)$ dimensional BPS Skyrme model \cite{BPS-star}. It also resembles the curve in the usual Skyrme model \cite{nelmes}. The end point is when $M_{ADM}=M^{max}_{ADM}$, above which the gravitating solitonic solution develops a singularity. 

\vspace*{0.2cm} 

As these $(3+1)$ dimensional gravitating Skyrmions are used to model neutron stars, this implies that our simplified $(2+1)$ dimensional theory can serve as a toy model where several nontrivial features of general neutron star solutions can be studied analytically. Some of these features are, e.g., the existence of a maximal mass beyond which the solution becomes singular, or the fact that for a sufficiently large mass the corresponding radius starts to diminish, or the exact calculation of the difference between nongravitating (proper) and total (ADM) mass. In view of our results, some of these properties can be analytically understood and estimated by the value of $\Omega$ or by formulas (\ref{DeltaR}) and (\ref{Rn_max}) and, therefore, can be directly related to particular target space integrals of the potential. 
%%%%%%%%%%%%%%%%%%%%%%%%%%%%%%%%%%%%%%%%%
\subsubsection{The new baby potential}
%%%%%%%%%%%%%%%%%%%%%%%%%%%%%%%%%%%%%%%%%
Another frequently used potential is the so-called new baby potential 
\be
\mathcal{U}= \frac{1}{4} h(1-h) \label{new-baby}
\ee
which has two minima at $h=0$ and $h=1$. The resulting BPS baby Skyrmions have ring like shapes with zero energy density at the origin. One can easily find that
\be
\langle \sqrt{\mathcal{U}} \rangle_{\mathbb{S}^2}= \frac{\pi}{16}, \;\;\; \left\langle \frac{1}{\sqrt{\mathcal{U}}} \right\rangle_{\mathbb{S}^2}=2\pi \;\;\; \mbox{and} \;\;\; \mathcal{A}= \frac{\pi^2}{16}
\ee
which means that $\Omega=2$. Quite surprisingly, we find that this potential gives {\it qualitatively} the same mass-radius curve as the maximally stiff, Heaviside theta potential. Of course, the slope of the $M$-$R^2$ curve can be different, as it depends of the ratio $\langle \mathcal{U}^{-1/2} \rangle_{\mathbb{S}^2} /(\langle \mathcal{U}^{1/2} \rangle_{\mathbb{S}^2} \Omega)$ and the dimensional parameters $\mu$ and $\kappa$.  However, multiplying the potential by a suitable numerical constant (which does not change $\Omega$) we can make that even the slopes coincide and the curves are {\it exactly identical}. 

In fact, there are infinitely many potentials leading to the $\Omega=2$ curve. One can show that all potentials which are invariant under the $h \rightarrow 1-h$ transformation have this property. The proof is as follows. We start with the definition of $\mathcal{A}$ and introduce the transformation $h\rightarrow 1-h$
\be
\mathcal{A}=\int_0^1\frac{dh}{\sqrt{\mathcal{U}(h)}} \int_h^1 \sqrt{\mathcal{U}(h')} dh' = \int_0^1  \frac{dh}{\sqrt{\mathcal{U}(h)}} \int_0^h \sqrt{\mathcal{U}(h')} dh'
\ee
where the last step follows from the invariance $\mathcal{U}(h)=\mathcal{U}(1-h)$. Hence,
\bea
\langle \mathcal{U}^{1/2} \rangle_{\mathbb{S}^2} \langle \mathcal{U}^{-1/2} \rangle_{\mathbb{S}^2} &=& \int_0^1\frac{dh}{\sqrt{\mathcal{U}(h)}} \int_0^1 \sqrt{\mathcal{U}(h')} dh'  \\
&=& \int_0^1\frac{dh}{\sqrt{\mathcal{U}(h)}}  \left(\int_0^h \sqrt{\mathcal{U}(h')} dh' + \int_h^1 \sqrt{\mathcal{U}(h')} dh' \right) = 2\mathcal{A}
\eea

This result shows that the relation between the shape of the potential and the mass-radius curve is more involved than one might expect. As a consequence, this makes the so-called inverse problem, i.e., the reconstruction of the equation of state from the $M$-$R$ curve, questionable. As we found, there can be quite many potentials (and therefore different equations of state) providing the same shape of the $M$-$R$ curve.  Furthermore, we proved that it is not just the shape of the potential which encodes the form of the mass-radius curve, but also its symmetries. 

%%%%%%%%%%%%%%%%%%%%%%%%%%%%%%%%%%%%%%%%%
\subsubsection{The old baby potential squared}
%%%%%%%%%%%%%%%%%%%%%%%%%%%%%%%%%%%%%%%%%
\begin{figure}
\hspace*{-1.0cm}
\includegraphics[height=5.cm]{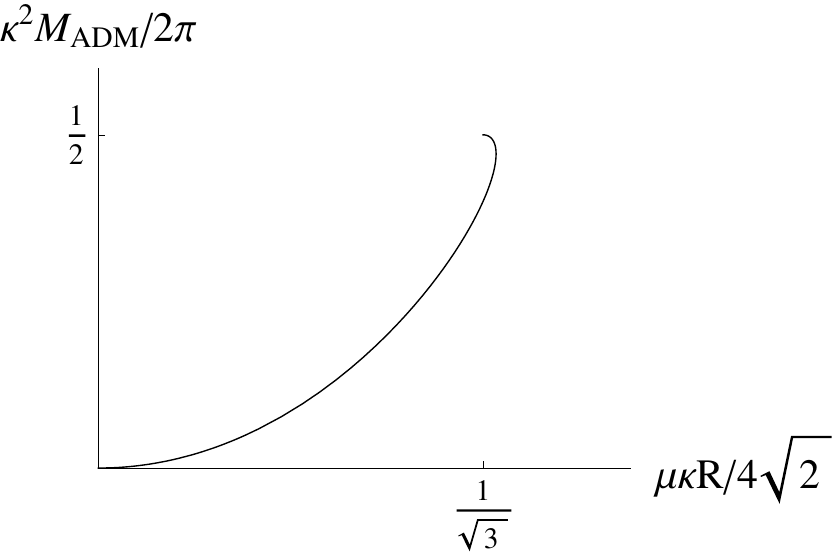}
\caption{Asymptotic mass-radius curve for the BPS baby Skyrmions with the old baby potential squared $\mathcal{U}=4\mathcal{U}_\pi^2$ in the asymptotically flat space-time.}
\label{MR-sqr}
\end{figure}
In the third example we consider the old baby potential squared 
\be
\mathcal{U}=4\mathcal{U}_\pi^2=\frac{h^2}{4} \label{old-sq} .
\ee
In contrast to the previous examples it supports infinitely extended, exponentially-like localised topological solitons
\be
h(z) = e^{-\frac{\mu z}{\lambda |n|}} .
\ee
This potential leads to the following averages 
\be
\langle \sqrt{\mathcal{U}} \rangle_{\mathbb{S}^2}= \frac{1}{4}, \;\;\; \left\langle \frac{1}{\sqrt{\mathcal{U}}} \right\rangle_{\mathbb{S}^2}=\infty .
\ee
Therefore, the ADM mass is 
\be
M_{ADM}=\frac{\pi}{4} \mu \lambda |n| \left( 1-\frac{1}{16} \kappa^2 \mu \lambda |n| \right)
\ee
with $n < n_{max}=\frac{8}{\kappa^2 \lambda \mu}$, while the geometric radius is of course infinite. To circumvent this problem we compute the usual mean square radius. This requires a knowledge of the metric function
\be
\bB^{-1/2}(z)= 1-\frac{\kappa^2 \lambda \mu |n|}{8} \left(1-e^{-\frac{2\mu z}{\lambda |n|} }  \right)
\ee
which is regular for $n<n_{max}$.  Hence, the radial coordinate $r$ reads
\be
\frac{r^2}{2} =  \int_0^z \bB^{-1/2}(z') dz' = \left( 1-\frac{\kappa^2 \lambda \mu |n|}{8} \right)z + \frac{\kappa^2 \lambda^2 n^2}{16} \left(1-e^{-\frac{2\mu z}{\lambda |n|}}  \right)
\ee
Now we can compute the mean square radius as 
\bea
\bar{R^2} &=& \frac{2\pi \int_0^\infty rdr r^2 \rho (r) }{2\pi \int_0^\infty rdr  \rho (r) } = \frac{2\pi}{M_{ADM}} \int_0^\infty rdr r^2 \rho (r) \\ 
&=&  \frac{4\pi}{M_{ADM}}  \int_0^\infty dz \left( \bB^{-1/2}(z) \rho(z) \int_0^z dz'  \bB^{-1/2}(z')  \right) 
\eea
And finally 
\be
\bar{R^2}=4\frac{\lambda}{\mu} |n| \frac{1-\frac{5}{32}\kappa^2 \lambda \mu |n|\left( 1 - \frac{1}{24} \kappa^2 \lambda \mu |n|\right)}{1-\frac{1}{16} \kappa^2\lambda \mu |n| }
\ee
or using $x=|n|/n_{max}$
\be
\bar{R^2}=\frac{32}{\kappa^2 \mu^2} x \frac{1-\frac{5x}{4} \left(1 -  \frac{x}{3} \right)}{1-\frac{x}{2} }
\ee
which is a finite function on the unit segment. The corresponding ADM mass-radius curve is plotted in Fig. \ref{MR-sqr}. As in the previous example with compact solitons, the curve is of the $\Omega<2$ type. It bends left for the topological charge close enough to $n_{max}$, which again coincides with the typical $(3+1)$ dimensional behaviour. 
%%%%%%%%%%%%%%%%%%%%%%%%%%%%%%%%%%%%%%%%%
\section{Cosmological constant  $\Lambda>0$ }
%%%%%%%%%%%%%%%%%%%%%%%%%%%%%%%%%%%%%%%%%
%%%%%%%%%%%%%%%%%%%%%%%%%%%%%%%%%%%%%%%%%
\subsection{Constant pressure (extremal) solutions}
%%%%%%%%%%%%%%%%%%%%%%%%%%%%%%%%%%%%%%%%%
Now we want to consider the BPS baby Skyrme model coupled with gravity in the presence of a positive cosmological constant. One can observe that equations  (\ref{LeB}), (\ref{LeA}), (\ref{LeM}) with $\Lambda >0$ possesses a formal solution 
\be
A=1 
\ee
\be
\kappa^2 p -2\Lambda=0  \;\;\; \Rightarrow \;\;\; p=\frac{2\Lambda}{\kappa^2} >0 \label{dSsolP}
\ee
which is nothing else but a non-zero pressure condition for the baby Skyrme matter. It represents a direct counterpart of the flat space zero-pressure (Bogomolny) equation.
Using the axially symmetric ansatz, the non-zero pressure condition gives
\be
 \frac{\lambda^2n^2}{\bB r^2} \frac{f^2f_r^2}{(1+f^2)^4} -\mu^2 \mathcal{U}(f) = \frac{2\Lambda}{\kappa^2}
\ee
which again can be cast into the non-gravity form using the variable $z$ (\ref{z})
\be
\lambda^2n^2 \frac{f^2f_z^2}{(1+f^2)^4} -\mu^2 \mathcal{U}(f) = \frac{2\Lambda}{\kappa^2}
\ee
or with the target space variable $h$ (\ref{h}) 
\be
 \frac{\lambda^2n^2}{4} h'^2- \mu^2 \mathcal{U}(h) =\frac{2\Lambda}{\kappa^2} . \label{LBPS}
\ee
This type of equation has been analyzed in connection with BPS Skyrmions under external positive pressure \cite{BPS-press}, which here is provided by the cosmological constant 
\be
p_{external}=\frac{2\Lambda}{\kappa^2}.
\ee
It is known that this equation does support topological solitons for any value of the pressure ($\Lambda>0$). However, these solitons are compactons i.e., with the vacuum approached at a finite distance (for {\it any} potential) and with a non-zero pressure at the boundary. It means that the corresponding solutions cannot be smoothly joined with the vacuum. Hence, such gravitating BPS baby Skyrmions make sense only if they fully cover de Sitter space-time, i.e., the size (volume) of the space is identical to the size $R$ of the baby Skyrmion. This gives a bound  on the radial coordinate of de Sitter space-time $r\in[0, r_{max}]$, where $r_{max}=R$. Equivalently, the metric should develop a singularity at the baby Skyrmion boundary. 
\\
To see this, we have to solve the matter equation {\it exactly}. First we observe that from eq. (\ref{LBPS})
(again the minus sign is chosen)
\be
-dz= \frac{\lambda |n|}{2} \frac{dh}{\sqrt{ \mu^2 \mathcal{U}(h) +\frac{2\Lambda}{\kappa^2}}} \label{dz-ds}
\ee
Hence,
\be
V=2\pi z_0 = \pi \frac{\lambda}{\mu} |n| \int_0^1 \frac{dh}{\sqrt{ \mathcal{U}(h) +\frac{2\Lambda}{\kappa^2\mu^2}}} =  \pi \frac{\lambda}{\mu} |n| \left\langle \frac{1}{\sqrt{ \mathcal{U}(h) +\frac{2\Lambda}{\kappa^2\mu^2}}}  \right\rangle_{\mathbb{S}^2}
\ee
The proper mass of the soliton can be also found in a geometric way. Namely, 
\be
M=\int d^2 x |g|^{\frac{1}{2}} \rho (\vec{x}) = 2\pi \int dz  \left( \frac{\lambda^2n^2}{4} h_z^2 +\mu^2 \mathcal{U}(h)\right)=2\pi \int dz  \left( 2\mu^2 \mathcal{U} + \frac{2\Lambda}{\kappa^2} \right)
\ee
where the last step follows from (\ref{LBPS}). Using now (\ref{dz-ds}) we get
\be
M=2\pi \lambda \mu |n| \int_0^1 \frac{\mathcal{U} + \frac{\Lambda}{\kappa^2\mu^2}}{\sqrt{\mathcal{U} + \frac{2\Lambda}{\kappa^2\mu^2}}} dh = 2\pi \lambda \mu |n|  \left\langle  \frac{ \mathcal{U} + \frac{\Lambda}{\kappa^2\mu^2}}{\sqrt{\mathcal{U} + \frac{2\Lambda}{\kappa^2\mu^2}}}  \right\rangle_{\mathbb{S}^2}.
\ee
Having solved the matter equation, we can consider the remaining equation for the metric function $\bB$. It is also exactly solvable in the $z$ variable
\be
\bB^{-1/2} (z)=1-\frac{\kappa^2}{2} \int_0^z \rho(h(z')) dz' - \Lambda z.
\ee
Now, our condition on this metric function differs from the flat space-time case. We require that the metric develops a singularity (horizon) at the boundary of the compact BPS baby Skyrmion, i.e., the soliton exists on the whole allowed space. This gives
\be
\frac{\kappa^2}{2} \int_0^{z_0} \rho(h(z')) dz' + \Lambda z_0 =1
\ee
which can be expressed by the global quantities as
\be
\frac{\kappa^2 M}{4\pi} +\frac{\Lambda V}{2\pi} =1 \label{extr}
\ee
This leads to a relation between the topological charge of the soliton, the parameters of the model and the value of the cosmological constant,
\be
\frac{1}{|n|} =  \frac{\kappa^2}{2} \lambda \mu \left\langle  \frac{ \mathcal{U} + \frac{\Lambda}{\kappa^2\mu^2}}{\sqrt{\mathcal{U} + \frac{2\Lambda}{\kappa^2\mu^2}}}  \right\rangle_{\mathbb{S}^2} +  \frac{\Lambda}{2} \frac{\lambda}{\mu}  \left\langle \frac{1}{\sqrt{ \mathcal{U}(h) +\frac{2\Lambda}{\kappa^2\mu^2}}}  \right\rangle_{\mathbb{S}^2}. \label{ds-extremal}
\ee
This formula is completely solution independent (except for the topological charge). 
The meaning of this relation is the following: if all parameters and the topological charge obey this formula, then the resulting gravitating soliton in the theory with positive cosmological constant is a constant pressure soliton given by the pertinent constant pressure generalisation of the BPS (zero pressure) equation. If the relation is not obeyed, then a solitonic solution may exist, however, it cannot be of the constant pressure (generalised BPS) type. 

Let us remark that such an extremal solution exists also for the pure quartic model without any potential term $\mathcal{U}=0$. Then,
\be
V=\frac{\pi}{\sqrt{2}} \frac{\lambda \kappa}{\sqrt{\Lambda} }|n|, \;\;\; M=\sqrt{2}\pi \frac{\lambda \sqrt{\Lambda}}{\kappa} |n|
\ee
with the extremal condition
\be
|n|=\frac{\sqrt{2}}{\lambda \sqrt{\Lambda} \kappa}
\ee
It is simply the \textit{dS} pressure which counterbalances the repulsion due to the four derivative term. 
%%%%%%%%%%%%%%%%%%%%%%%%%%%%%%%%%%%%%%%%%
\subsection{No-backreaction approximation}
%%%%%%%%%%%%%%%%%%%%%%%%%%%%%%%%%%%%%%%%%
To attack the problem of the existence of non-extremal gravitating solitons, we change the metric to a more convenient form
\be
ds^2=e^{-2 \bd (r)} \ba (r) dt^2 - \frac{1}{\ba (r)} dr^2 -r^2 d\varphi^2 .
\ee
Then, we get
\eqn
\ba' & = & -\kappa^2 r \left(\rho+\frac{2\Lambda}{\kappa^2} \right)=  -\kappa^2 r \left(\frac{\ba \lambda^2 n^2}{4r^2} h'^2+\mu^2\mathcal{U}+\frac{2\Lambda}{\kappa^2} \right)\label{Lea}\\
\delta' & = & -\frac{\kappa^2 r}{2\ba}  \left( \rho + p \right)  =  -\frac{\kappa^2 r}{2\ba}  \left( \frac{2\ba \lambda^2 n^2}{4r^2} h'^2\right)=  -  \frac{\kappa^2  \lambda^2 n^2}{4r} h'^2 \label{Led} \\
0&=&\partial_r \left( \frac{1}{r} e^{-\delta} \ba h' \right) - \frac{2\mu^2}{\lambda^2 n^2} re^{-\delta} \mathcal{U}_h. \label{Leh}
\eqnx
First of all, this set can be simplified if we introduce $y=r^2/2$
\eqn
\ba_y & = &  -\kappa^2  \left(\frac{\ba \lambda^2 n^2}{4} h_y^2+\mu^2\mathcal{U}+\frac{2\Lambda}{\kappa^2} \right)\label{yLea}\\
\delta_y & = & -  \frac{\kappa^2  \lambda^2 n^2}{4} h_y^2 \label{yLed} \\
0&=&\partial_y \left( e^{-\delta} \ba h_y \right) - \frac{2\mu^2}{\lambda^2 n^2} e^{-\delta} \mathcal{U}_h. \label{yLeh}
\eqnx
Next, the second equation can be used to get rid of the function $\delta$ from the last equation
\be
e^{-\delta}  \partial_y \left( \ba h_y \right) - e^{-\delta} \delta_y \ba h_y - \frac{2\mu^2}{\lambda^2 n^2} e^{-\delta} \mathcal{U}_h =0
\ee
i.e.,
\be
  \partial_y \left( \ba h_y \right) +  \frac{\kappa^2  \lambda^2 n^2}{4} h_y^3 \ba - \frac{2\mu^2}{\lambda^2 n^2}  \mathcal{U}_h =0.
\ee
Then,
\eqn
  0&=&\ba h_{yy} +\ba_y h_y +  \frac{\kappa^2  \lambda^2 n^2}{4} h_y^3 \ba - \frac{2\mu^2}{\lambda^2 n^2}  \mathcal{U}_h \nonumber \\
  & =&\ba h_{yy} -\kappa^2  \left(\frac{\ba \lambda^2 n^2}{4} h_y^2+\mu^2\mathcal{U}+\frac{2\Lambda}{\kappa^2} \right) h_y +  \frac{\kappa^2  \lambda^2 n^2}{4} h_y^3 \ba - \frac{2\mu^2}{\lambda^2 n^2}  \mathcal{U}_h \nonumber \\
  & =&\ba h_{yy} -\kappa^2  \left(\mu^2\mathcal{U}+\frac{2\Lambda}{\kappa^2} \right) h_y  - \frac{2\mu^2}{\lambda^2 n^2}  \mathcal{U}_h . \label{full h}
\eqnx
Here we begin with the no-backreaction approximation, which means that $\kappa=0$. Then, the field equations are solved by
\be
\ba = 1-2\Lambda y, \;\;\; \delta=0
\ee
where the radial coordinate $y \in \left[0, \frac{1}{2\Lambda}\right]$. Hence, we are left only with the second order matter equation
\be
\partial_y \left((1-2\Lambda y) h_y \right) - \frac{2\mu^2}{\lambda^2 n^2} \mathcal{U}_h =0.
\ee
This equation cannot be integrated to a Bogomolnyi type equation. However, we can solve it exactly for the old baby potential (\ref{old}). 
Indeed, we get
\be
\partial_y \left((1-2\Lambda y) h_y \right) - \frac{\mu^2}{2\lambda^2 n^2} =0.
\ee
Thus,
\be
(1-2\Lambda y) h_y =  \alpha y +C_0, \;\;\;\; \alpha \equiv \frac{\mu^2}{2\lambda^2 n^2}
\ee
that is
\be
 h_y =   \frac{C_0+\alpha y}{1-2\Lambda y} .
\ee
Hence, a general solution is
\be
h=A-\frac{\alpha}{2\Lambda} y -\frac{1}{2\Lambda} \left( \frac{\alpha}{2\Lambda} +C_0\right) \ln (1-2\Lambda y)
\ee
where one has to remember that $y \leq \frac{1}{2\Lambda}$. 
After imposing the boundary conditions
\be
h(y=0)=1, \;\; h(y=Y)=0, \;\;\; h_y(y=Y)=0
\ee
we find 
\be
h=1-\frac{\alpha}{(2\Lambda)^2}\left[ 2\Lambda y + \left(1- 2\Lambda Y \right) \ln (1-2\Lambda  y)\right]
\ee
where the size of the compact soliton $Y$ obeys
\be
\frac{4\Lambda^2}{\alpha} =(1-2\Lambda Y) \ln (1-2\Lambda Y) +2\Lambda Y \label{dS-Y}
\ee
After eliminating $\alpha$, we get the following form of the solution parametrised by its size $Y$
\be
h=1- \frac{ 2\Lambda y + \left(1- 2\Lambda Y \right) \ln (1-2\Lambda  y)}{2\Lambda Y + \left(1- 2\Lambda Y \right) \ln (1-2\Lambda  Y)} \label{sol dS ext}
\ee
where $y \in [0,Y]$. As the size of the soliton cannot exceed the size of the space-time we get a condition on the parameters of the model and the topological charge of the solution. Namely, the right hand side of (\ref{dS-Y}) is a positive function bounded by 1 for $Y=1/(2\Lambda)$. Hence, $4\Lambda^2 < \alpha$ i.e., 
\be
 |n| \leq \frac{\mu}{2\sqrt{2} \lambda \Lambda}. \label{cond-n-ds}
\ee
This is the maximal charge of the axially symmetric solution which exists for a given $\Lambda$ in the no-backreaction approximation.  The total energy of such a configuration is
\be
E= \frac{\pi}{8} \frac{\mu^2}{2\Lambda} \left[2+4\Lambda Y -\frac{(2\Lambda Y)^2}{2\Lambda Y + \left(1- 2\Lambda Y \right) \ln (1-2\Lambda  Y)}  \right] .
\ee

Note that one can get the limiting no-backreaction solution which occupies the whole \textit{dS} space-time, i.e., when $Y=1/(2\Lambda)$. It is simply given by
\be
h=1-2\Lambda y, \;\;\; y \leq \frac{1}{2\Lambda} \label{ds-sol-lim}
\ee
and $ |n| = \frac{\mu}{2\sqrt{2} \lambda \Lambda}$. In contrast to the other solutions, this solution does not obey the boundary condition $h_y(Y)=0$ because $h_y(y=1/(2\Lambda)) = -2\Lambda \neq 0$. Therefore, there is a strong qualitative difference between the smooth solutions (\ref{sol dS ext}) and the limiting solution (\ref{ds-sol-lim}), which resembles a first order phase transition. 

Finally, when the parameters do not obey condition (\ref{cond-n-ds}), there are no topologically non-trivial solutions at all. This differs for example from the BPS baby Skyrme model on a two dimensional sphere, where for sufficiently large topological charge compact solitons (obeying a Bogomony equation) are replaced by solutions fully covering the (compact) base space which do not obey the Bogomolny equation but still solve the second-order field equation. There is a qualitative change of the type of solution, but baby Skyrmions do exist for any topological charge. Here, in the \textit{dS} background case, solutions cannot exist beyond a maximal topological charge. 
%%%%%%%%%%%%%%%%%%%%%%%%%%%%%%%%%%%%%%%%%
\subsection{Fully gravitating system}
%%%%%%%%%%%%%%%%%%%%%%%%%%%%%%%%%%%%%%%%%
\begin{figure}
\hspace*{-1.0cm}
\includegraphics[height=5.cm]{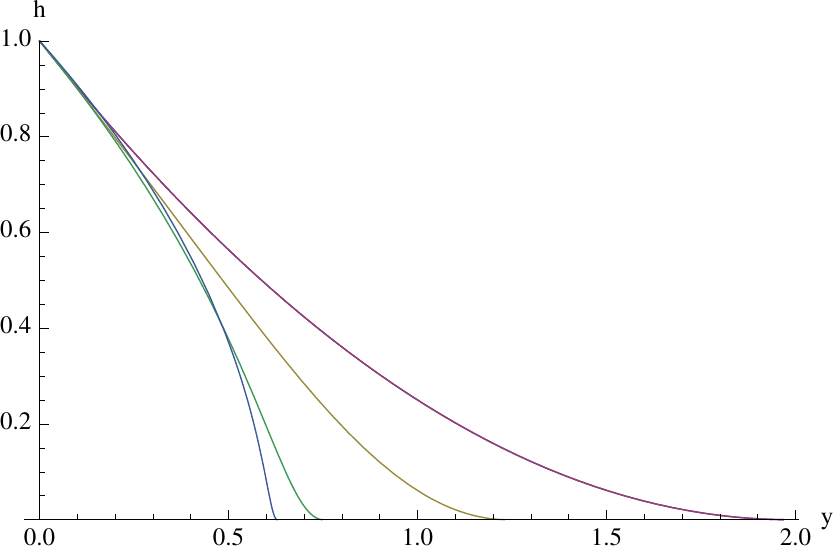}
\caption{Profile function of charge one BPS baby Skyrmions for the the baby potential $\mathcal{U}_\pi$ with $\mu=\lambda=1$ and $\Lambda=1$, for different values of the gravitational constant: $\kappa^2=0, \kappa^2=3, \kappa^2=5$ and $\kappa^2=5.89$ which is the extremal limit - lines: red, yellow, green and blue respectively.}
\label{ds}
\end{figure}

Here we want to find non-extremal solutions in the full gravitating system, for which the local pressure is no longer constant. As an example, we consider the old baby potential. 

In Fig. \ref{ds} we show the profile function $h$ for the charge one solution and different values of the gravitational constant $\kappa^2$. The model parameters are chosen as $\Lambda=0.01$ and $\mu=\lambda=1$. We begin with a vanishing gravitational constant, which is just the above considered no-backreaction approximation, and get a compacton in the de Sitter background. The values of the constants obey condition (\ref{cond-n-ds}) and a charge one solution does exist. It is given by formula (\ref{sol dS ext}) with the compacton size $Y=1.987$, which perfectly agrees with our numerics. The size of the de Sitter space time is $1/(2\Lambda)=50$. Hence, the compacton is much smaller that the allowed space. 

Once we take a non-zero $\kappa$ then we can study the impact of matter on space-time. An overall effect is that increasing $\kappa$ leads to a shrinking of the compacton. At the same time, the size of the de-Sitter space-time also decreases, but we still find a compacton with a smooth (differentiable) approach to the vacuum. However, this approach occurs in a more and more rapid fashion as the compacton starts to feel the reduced size of the space-time. At some point the size of the space-time reaches the size of the compacton. This happens approximately for $\kappa^2=5.8$, which clearly coincides with the extremal solution which occurs for $\kappa^2=5.889$.

%%%%%%%%%%%%%%%%%%%%%%%%%%%%%%%%%%%%%%%%%
\section{Asymptotic $AdS$ space-time}
%%%%%%%%%%%%%%%%%%%%%%%%%%%%%%%%%%%%%%%%%
%%%%%%%%%%%%%%%%%%%%%%%%%%%%%%%%%%%%%%%%%
\subsection{Non-existence of  extremal solutions}
%%%%%%%%%%%%%%%%%%%%%%%%%%%%%%%%%%%%%%%%%
Let us now turn to the asymptotic \textit{AdS} space-time, which is probably the most interesting and physically important case, owing to its relation to holographic toy models of dense nuclear matter/QCD. As we remarked in the Introduction, there is a $(2+1)$ dimensional analogue of the Sakai-Sugimoto holographic description of baryons which, in its simplest set-up, is given by the baby Skyrme model on an \textit{AdS} background. However, since the $O(3)$ $\sigma$-model part of the full action does not support topological solitons for negative cosmological constant, it is the BPS part (quartic term plus the potential) which must trigger such baby Skyrmions. Therefore, an understanding of the properties of  baby Skyrmions in the BPS Skyrme model on \textit{AdS} space is the first step in an analytical understanding of properties of dense baryonic matter in a holographic approach, at least in its simplified $(2+1)$ dimensional version. 

We consider equations (\ref{LeB}), (\ref{LeA}), (\ref{LeM}) with $\Lambda<0$. Similarly as in the de Sitter case, there is a solution 
\be
A=1 
\ee
\be
\kappa^2 p -2\Lambda=0. \label{LsolP}
\ee 
After insertion of the axially symmetric ansatz, the non-zero pressure condition gives
\be
 \frac{\lambda^2n^2}{\bB r^2} \frac{f^2f_r^2}{(1+f^2)^4} -\mu^2 \mathcal{U}(f) = \frac{2\Lambda}{\kappa^2}
\ee
which once again can be brought to a non-gravity form using the variable $z$, (\ref{z})
\be
\lambda^2n^2 \frac{f^2f_z^2}{(1+f^2)^4} -\mu^2 \mathcal{U}(f) = \frac{2\Lambda}{\kappa^2}.
\ee
Unfortunately, there is no topologically nontrivial solutions of this equation for negative $\Lambda$. Indeed, it is not possible for the function $f$ to reach the the vacuum value $f=0$. At this point the potential $\mathcal{U}$ vanishes, and the left hand side takes a nonnegative value. However, it has to be equal to $2\Lambda/\kappa^2$ which is a negative constant. Hence, $f$ cannot reach the vacuum. 

Of course, this means that topological solutions cannot be of this simple form which results in the observation that the BPS property of the gravitating BPS baby Skyrme model is lost when a nonzero negative cosmological constant is added.  Whether there are gravitating solitons obeying a more complicated matter equation is a different question. 

%%%%%%%%%%%%%%%%%%%%%%%%%%%%%%%%%%%%%%%%%
\subsection{No-backreaction approximation}
%%%%%%%%%%%%%%%%%%%%%%%%%%%%%%%%%%%%%%%%%
It is interesting that, although the model loses completely its BPS property, it is {\it solvable} in an \textit{AdS} background for two potentials which are widely  considered in physical applications of the baby Skyrme model. 
%%%%%%%%%%%%%%%%%%%%%%%%%%%%%%%%%%%%%%%%%
\subsubsection{The old baby potential $\mathcal{U}_\pi$ }
%%%%%%%%%%%%%%%%%%%%%%%%%%%%%%%%%%%%%%%%%
The first {\it analytically} solvable example is, again, provided by the old baby potential (\ref{old}). Then, we can adopt the previously found solution in de Sitter space, taking $\Lambda <0$
\be
h=1- \frac{ 2\Lambda y + \left(1- 2\Lambda Y \right) \ln (1-2\Lambda  y)}{2\Lambda Y + \left(1- 2\Lambda Y \right) \ln (1-2\Lambda  Y)} \label{h-ads}
\ee
where the size of the compact soliton $Y$ reads
\be
\frac{(-2\Lambda)^2}{\alpha} =(1-2\Lambda Y) \ln (1-2\Lambda Y) +2\Lambda Y \label{Y-ads}
\ee
Now $y\in [0, \infty]$  and there is no restriction on $Y$ and, therefore, no bounds on the topological charge. Equivalently, the right hand side of (\ref{Y-ads}) is a monotonously growing function of $-2\Lambda Y$ from 0 to infinity. 
This solution leads to the following energy expression
\be
E= \frac{\pi}{8} \frac{\mu^2}{2|\Lambda|} \left[- 2+4|\Lambda| Y +\frac{(2\Lambda Y)^2}{-2|\Lambda| Y + \left(1+ 2|\Lambda| Y \right) \ln (1+2|\Lambda|  Y)}  \right] . \label{MR-kappa0}
\ee
An important question is whether multi-soliton solutions are stable against decay into smaller charge constituents. First, we consider a small cosmological constant limit where $\frac{\Lambda^2}{\alpha} \ll 1$. This means that the cosmological constant is much smaller than the inverse of the characteristic length scale of the BPS baby Skyrmion, $\Lambda \ll \frac{\mu}{\lambda}$. Then the size of the soliton is 
\be
Y =  \frac{1}{|\Lambda|} \left( 2\frac{\lambda |\Lambda|}{\mu} |n| + \frac{4}{3}  \frac{\lambda^2 \Lambda^2}{\mu^2} n^2 \right) +o(\Lambda^2 \alpha^{-1})
\ee
while the energy is
\be
E=\frac{\mu^2}{|\Lambda|} \left( \frac{2\pi}{3} \frac{\lambda |\Lambda|}{\mu} |n| + \frac{2\pi}{9} \frac{\lambda^2 \Lambda^2}{\mu^2}  n^2 \right)+o(\Lambda^2 \alpha^{-1}) ,
\ee
where the first terms are, of course, the size and energy in asymptotically flat space $\Lambda =0$. Then, 
\be
E(n=2)-2E(n=1)= 2 \frac{2\pi}{9} \lambda^2 |\Lambda| +o(\Lambda^2 \alpha^{-1}) >0
\ee
and the charge two axially symmetric soliton is heavier than two charge one solitons. The same happens in the large cosmological constant limit $\frac{\Lambda^2}{\alpha} \gg1$, that is $\Lambda \gg \frac{\mu}{\lambda}$. It means that the left hand side of (\ref{Y-ads}) tends to infinity. The same must happen for the quantity $-2\Lambda Y$. Hence,
\be
\frac{8\lambda^2 \Lambda^2}{\mu^2} n^2 \approx 2|\Lambda| Y \ln (2 |\Lambda|Y )
\ee
This implies that the size $Y$ grows more than linearly (almost quadratically) with $\frac{2\sqrt{2} \lambda |\Lambda|}{\mu} |n|$. Now, the energy 
\be
E\approx \frac{\pi}{4} \mu^2 Y
\ee
also grows more than linearly with $\frac{2\sqrt{2} \lambda |\Lambda|}{\mu} |n|$. Consequently, it grows more than linearly with the topological charge. This again implies that $E(n=2)-2E(n=1) >0$. 

\vspace*{0.2cm}

The message we can learn from this result is that the energy of the charge two soliton is bigger that twice the energy of the charge one soliton. However, this does not necessary mean that our exact axially symmetric solutions are unstable towards decay into two charge one compact constituents. This is related to the fact that AdS space-time is not translationally invariant. In other words, moving a soliton from the point $r=0$  to an arbitrary position in AdS requires an additional amount of energy. In fact, the impact of AdS space-time on solitons can be modeled by putting solitons in flat space with an external position dependent potential which generates a force (pressure) moving the soliton towards the origin. 

Hence, at this stage of our investigations we may conclude that we found multi-soliton solutions which resemble the {\it liquid phase} previously found in the flat space. Here, however, all volume preserving diffeomorphism symmetries of the static solutions are lost, because of the nature of the background space-time. 
It is an interesting question whether there exist other phases consisting of identifiable charge one substructures (e.g. a crystal of gaseous phase), corresponding to the gaseous phase of the flat space model. 
To answer this question and to figure out which phase provides the lower energy state, however, requires full two dimensional computations, which are beyond the scope of this paper.

%%%%%%%%%%%%%%%%%%%%%%%%%%%%%%%%%%%%%%%%%
\subsubsection{The old baby potential squared $\mathcal{U}=4\mathcal{U}_\pi^2$} 
%%%%%%%%%%%%%%%%%%%%%%%%%%%%%%%%%%%%%%%%%
Now we turn our attention to the old baby potential squared (\ref{old-sq}). 
Then, we get
\be
\partial_y \left((1-2\Lambda y) h_y \right) - \frac{\mu^2}{\lambda^2 n^2} h=0
\ee
It is convenient to introduce a new variable $z=1-2\Lambda y$. Now,
\be
\partial_z \left(z h_z \right) -\beta^2 h=0 \;\;\; \Rightarrow \;\;\; zh_{zz}+h_z-\beta^2h=0
\ee
where
\be
\beta^2=\frac{\mu^2 }{4\lambda^2 \Lambda^2n^2} .
\ee
This equation has a general solution given by the modified Bessel functions 
\be
h(z)=c_1 I_0(2\beta\sqrt{z}) +c_2K_0(2\beta \sqrt{z})
\ee
or 
\be
h(y)=c_1 I_0(2\beta\sqrt{1-2\Lambda y}) +c_2K_0(2\beta \sqrt{1-2\Lambda y}).\label{h-squared}
\ee
Imposing the boundary conditions $h(0)=1$ and $h(\infty)=0$ we find the exact solution
\be
h(y)=\frac{1}{K_0\left(2\beta \right)} K_0 \left(2\beta \sqrt{1-2\Lambda y}\right) .
\ee
Now, the energy reads
\be
E=2 \pi  \int_0^\infty dy \left((1-2\Lambda y) \frac{\lambda^2n^2}{4} h_y^2 +\frac{\mu^2}{4}h^2 \right) .
\ee
Inserting our solution we get
\be 
E= \frac{\pi}{4} \frac{\mu^2 }{-\Lambda} \left(\frac{K_2(2\beta)}{K_0(2\beta)}-1\right)  \label{e-sq-ads}
\ee
where we used that $K'_0(z)=-K_1(z)$ and
$$ \int xK_0^2(x) dx = \frac{1}{2}x^2(K_0^2(x)-K_1^2(x)) \;\;\; \mbox{ and } \;\;\; \int xK_1^2(x) dx = \frac{1}{2}x^2(K_1^2(x)-K_0(x) K_2(x)).$$
In the limit of vanishing cosmological constant we get the energy in flat space-time $E=\frac{\pi}{2}\mu \lambda |n|$. 
One can also show that 
\be
E(n=2)-2E(n=1)= \frac{\pi}{4} \mu \lambda\;  b \left(\frac{K_2(b/2)}{K_0(b/2)}-2\frac{K_2(b)}{K_0(b)} +1\right)  >0 \label{be-ads}
\ee
where
\be
b=n2\beta = \frac{\mu }{-  \lambda \Lambda} .
\ee
This means that the axially symmetric charge $n=2$ solution is heavier than twice the axially symmetric charge one solution, for any values of the parameters. 

It is worth noticing that the excess of energy (\ref{be-ads}) goes to 0 for $b \rightarrow \infty$. As expected, it corresponds to the flat space-time limit $\Lambda \rightarrow 0$. 

%%%%%%%%%%%%%%%%%%%%%%%%%%%%%%%%%%%%%%%%%
\subsection{Fully gravitating system}
%%%%%%%%%%%%%%%%%%%%%%%%%%%%%%%%%%%%%%%%%
\begin{figure}
\hspace*{-1.0cm}
\includegraphics[height=5.cm]{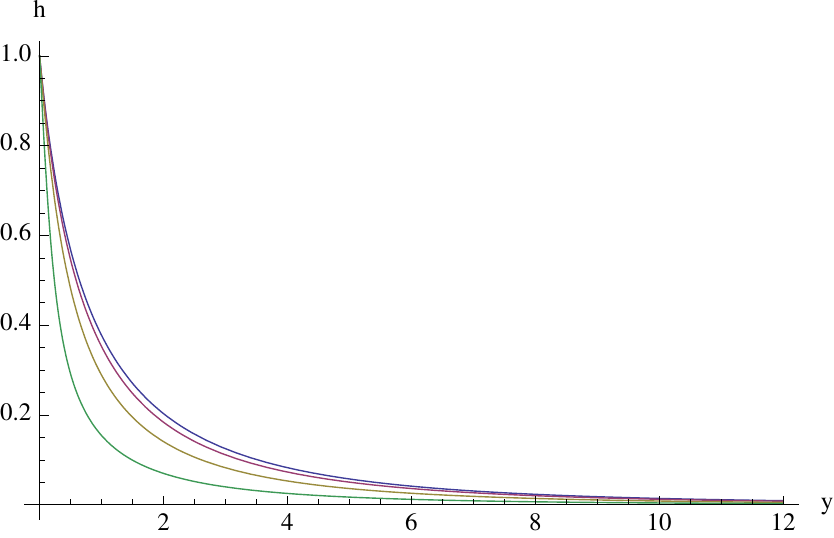}
\includegraphics[height=5.cm]{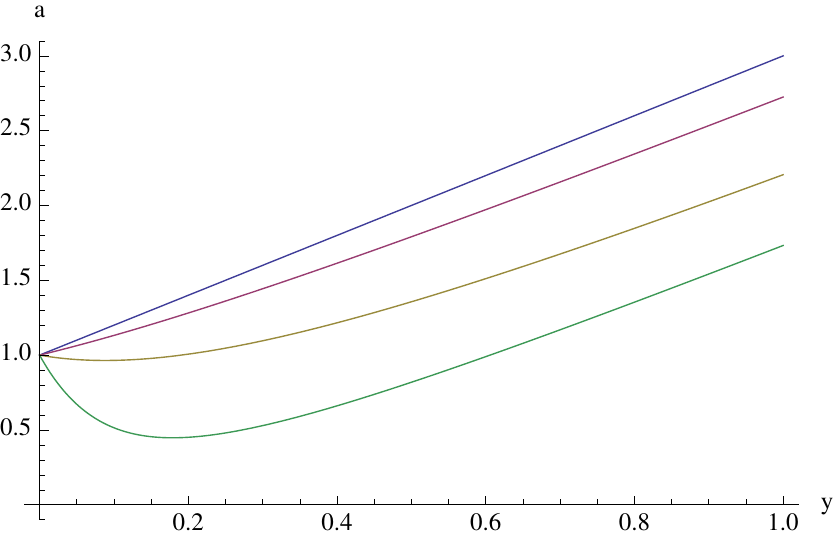}
\caption{Charge one BPS baby Skyrmions for the baby potential squared $\mathcal{U}=2\mathcal{U}_\pi^2$ with $\mu=\lambda=1$ and $\Lambda=-1$ for different values of gravity constant ($\kappa^2=0, \kappa^2=1, \kappa^2=3$ and maximal $\kappa^2=4.985$ - lines: blue, red, yellow and green respectively). Left panel: matter profile function $h$. Right panel: metric function $\ba$.}
\label{U2}
\end{figure}
\begin{figure}
\hspace*{-1.0cm}
\includegraphics[height=5.cm]{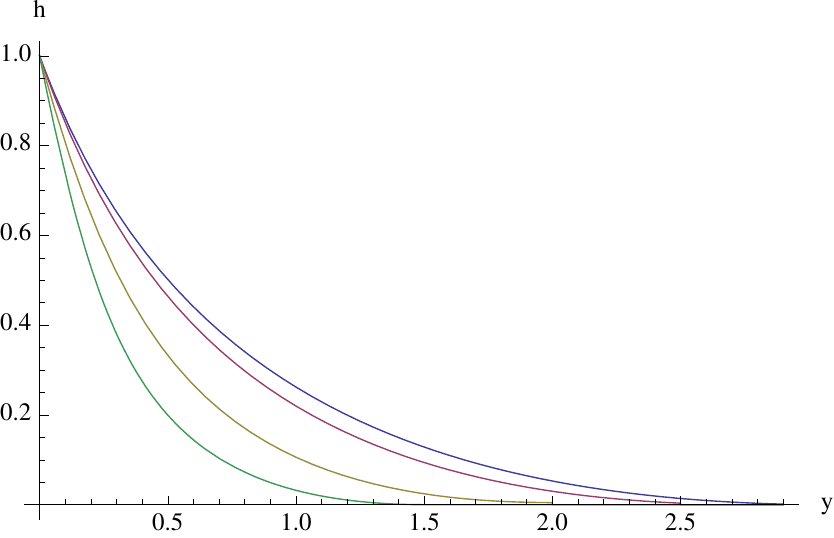}
\caption{Profile function of charge one BPS baby Skyrmions for the baby potential $\mathcal{U}_\pi$ with $\mu=\lambda=1$ and $\Lambda=-1$ for different values of gravity constant: $\kappa^2=0, \kappa^2=1, \kappa^2=2$ and maximal $\kappa^2=3.65$ - lines: blue, red, yellow and green respectively.}
\label{U1}
\end{figure}
\begin{figure}
%\hspace*{-1.0cm}
\includegraphics[height=4.5cm]{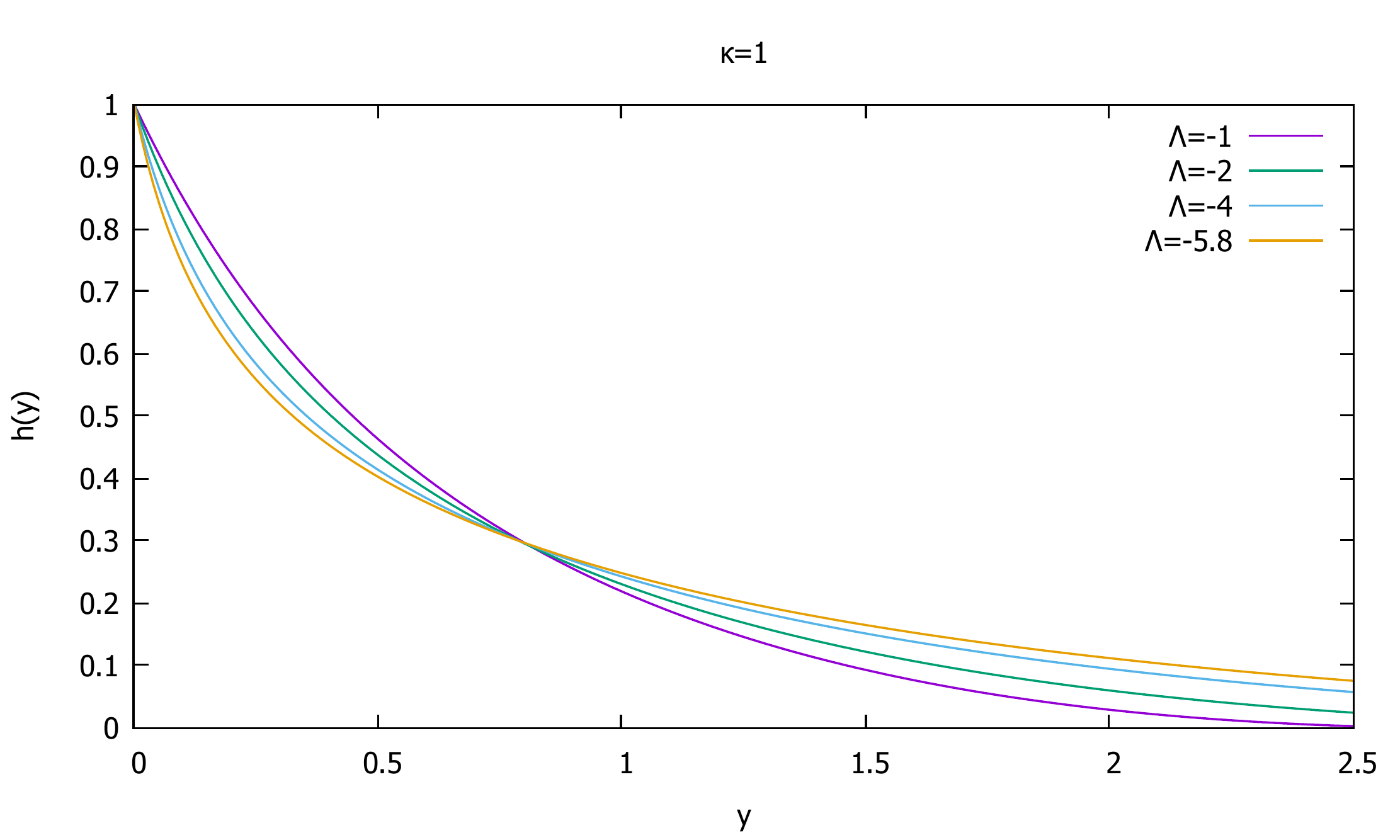}
\includegraphics[height=4.5cm]{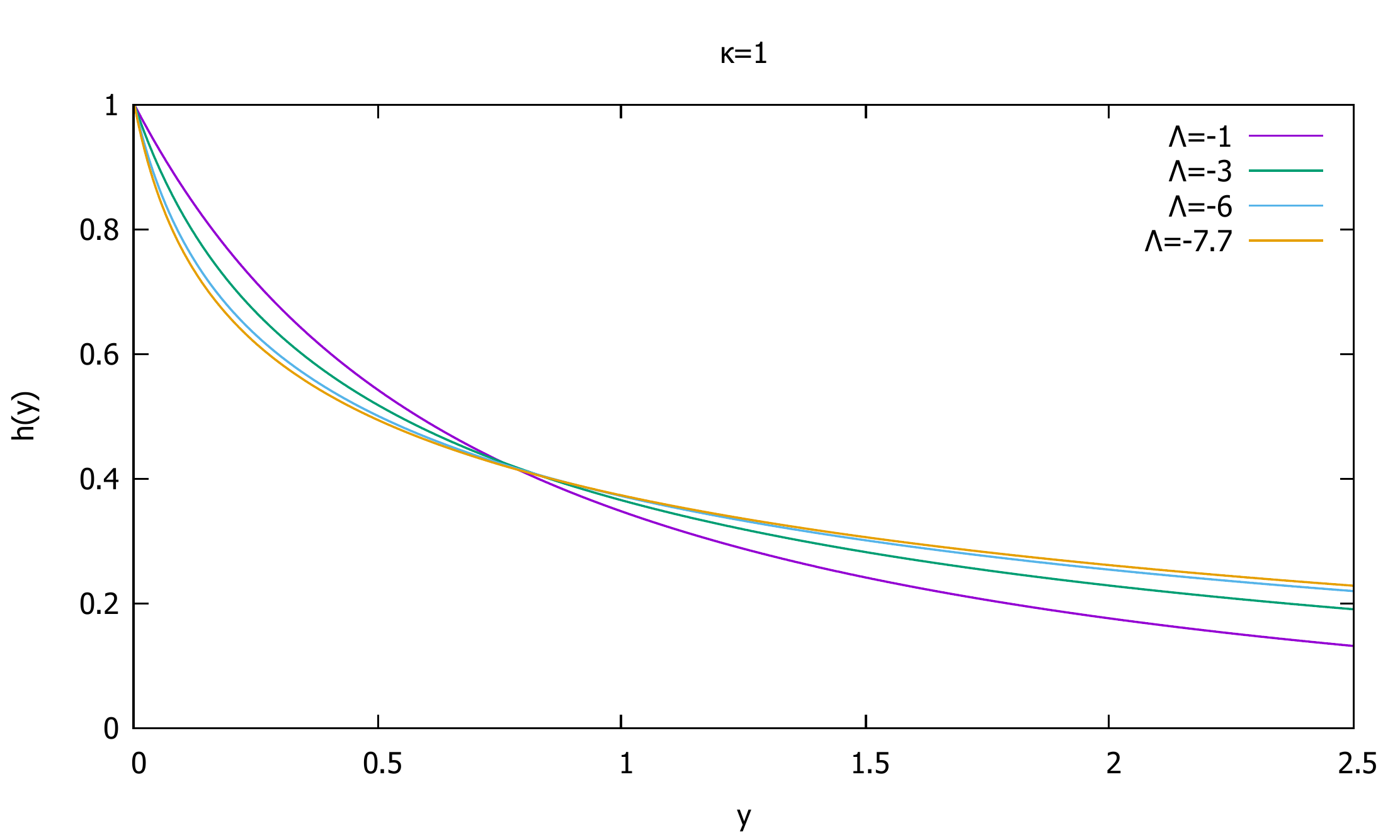}
\includegraphics[height=4.5cm]{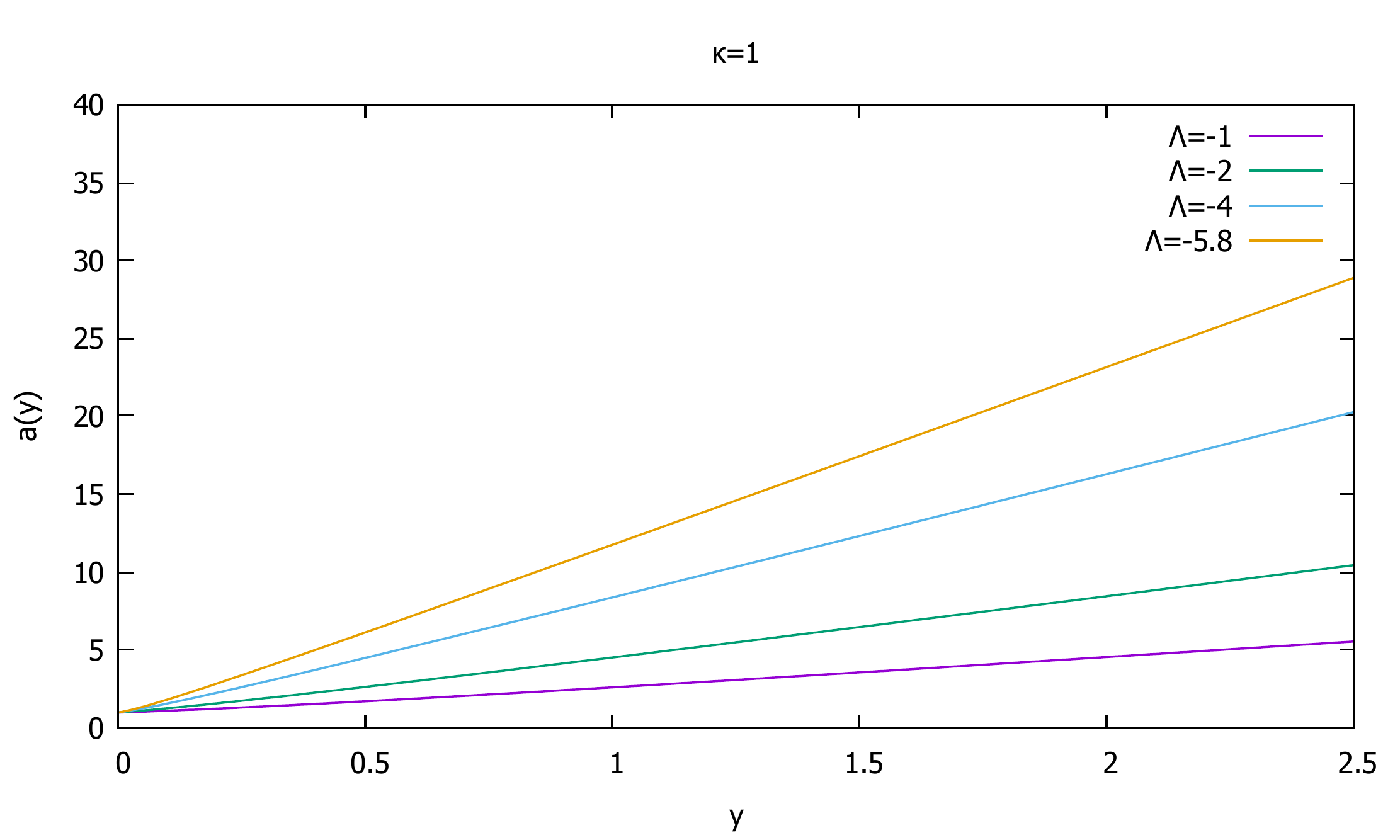}
\includegraphics[height=4.5cm]{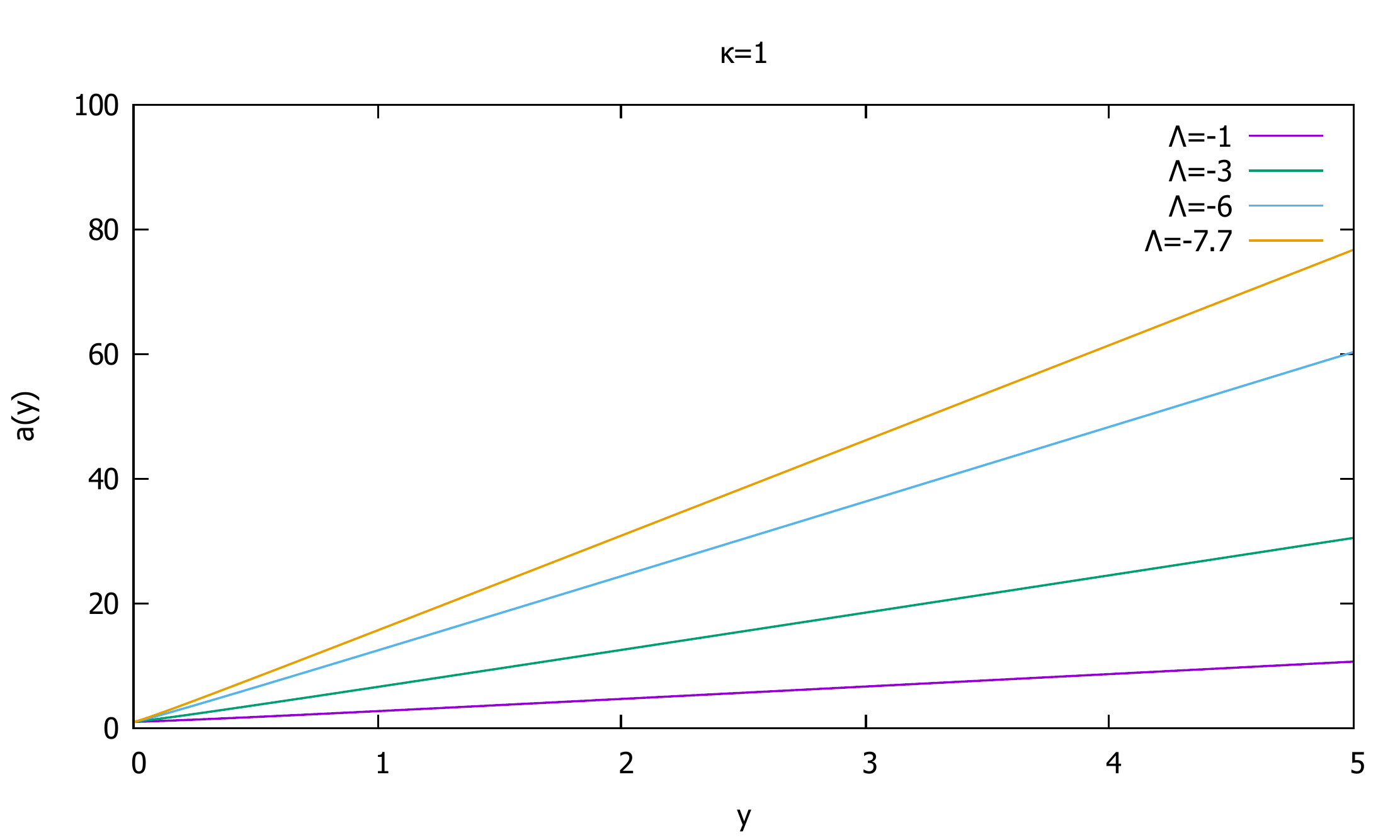}
\caption{Profile function $h$ (upper) and metric function $\ba$ (lower) of charge one BPS baby Skyrmions for $\mathcal{U}_\pi$ (left) and  $2\mathcal{U}_\pi^2$ (right) with $\kappa=1$ and for different $\Lambda$.}
\label{L}
\end{figure}

Finally, we solve the gravitating BPS baby Skyrme model numerically with the assumption of the axial symmetry (\ref{yLea}, \ref{full h}). For concreteness we analyse two potentials for which we have analytical solutions in $\kappa=0$ limit, i.e., the old baby and the old baby squared. 

In Fig. \ref{U2} we plot charge one solitonic solutions for the old baby potential squared $\mathcal{U}=2\mathcal{U}_\pi^2$. Here we choose for the matter coupling constants the values $\mu =\lambda=1$, while $\Lambda=-1$. Then, we vary the gravity coupling constant $\kappa^2$. For $\kappa^2 \rightarrow 0$, the solution smoothly tends to the no-backreaction approximation. On the other hand, for growing $\kappa^2$, the matter field  is squeezed while the metric function $\ba$ slowly changes its character - from a monotonously increasing function to a function which possesses a local minimum. Of course, asymptotically $\ba \rightarrow \ba_\infty=2y$. The maximal value of the gravity constant is $\kappa^2=4.985$. Above this value, there are no topologically nontrivial solutions as it is impossible for the profile function $h$ to reach its vacuum value $h=0$. 

In Fig. \ref{U1} we also show the profile function of the charge one BPS baby Skyrmions for the old baby potential, for different $\kappa$. The size of the compacton gets smaller with growing gravity coupling constant, until it reaches the maximum value $\kappa^2=3.65$ beyond which no topological solution exists. 
 
 In Fig. \ref{L} we present the profile function of the baby Skyrmion as well as the metric function $\ba$ for the old baby and old baby potential squared with $\mu=\lambda=\kappa=1$ while we change $\Lambda$. The impact of $\Lambda$ on the profile function is twofold. In the region near the origin, where most of the mass is concentrated, increasing the cosmological constant leads to a more rapid decrease of $h$. In the asymptotic region, where the solution approaches the vacuum, we observe the opposite behavior. The profile function decreases  slower for bigger $\Lambda$, which is clearly related to the negative pressure-like behaviour of the negative cosmological constant.  
 
\begin{figure}
%\hspace*{-1.0cm}
\includegraphics[height=5.5cm]{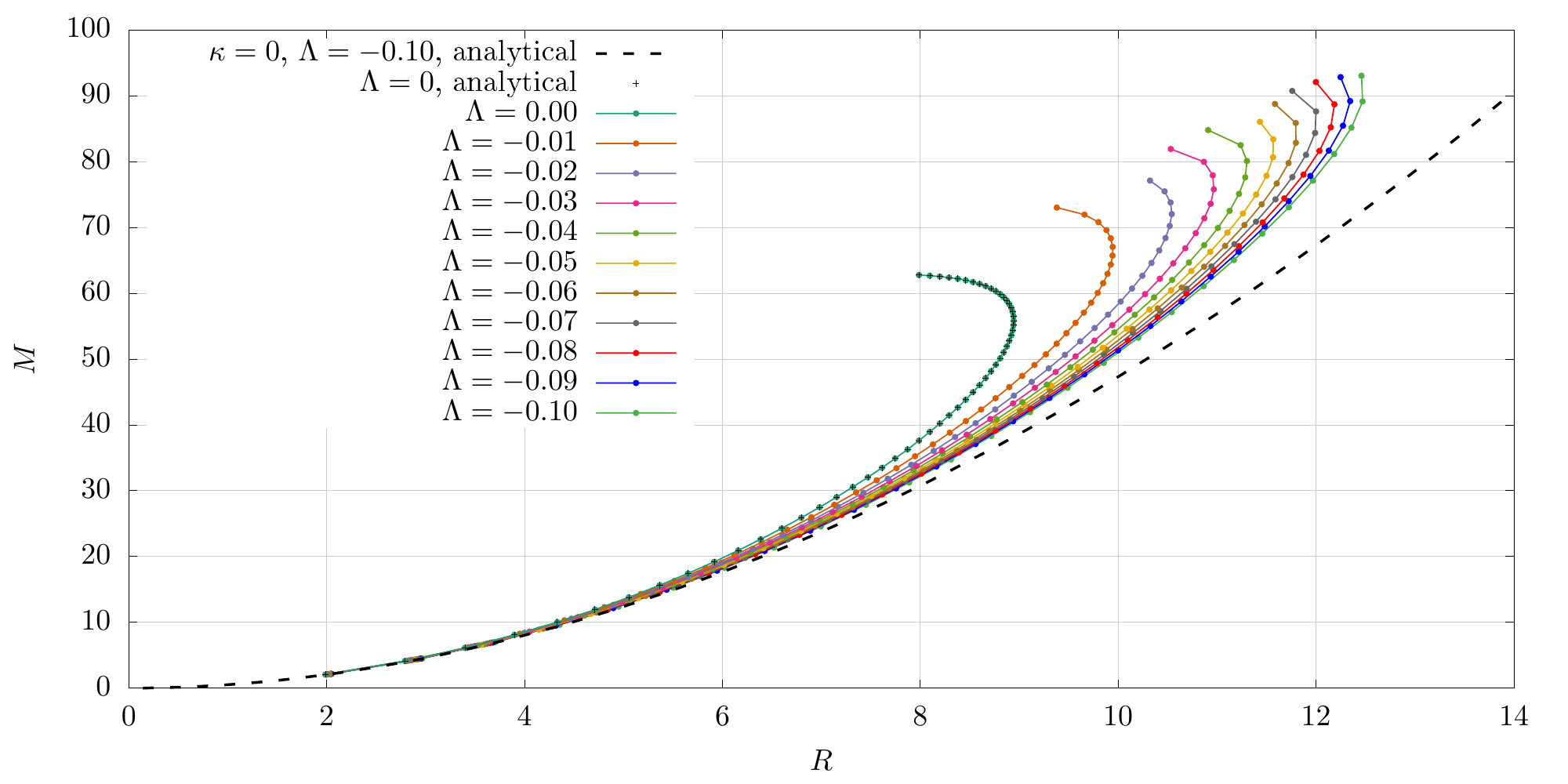}
\caption{The mass-radius curve for the old baby potential with $\lambda=\mu=1$ and $\kappa^2=0.1$, for different $\Lambda$. }
\label{MR-AdS-1}
\end{figure}

\begin{figure}
%\hspace*{-1.0cm}
\includegraphics[height=5.5cm]{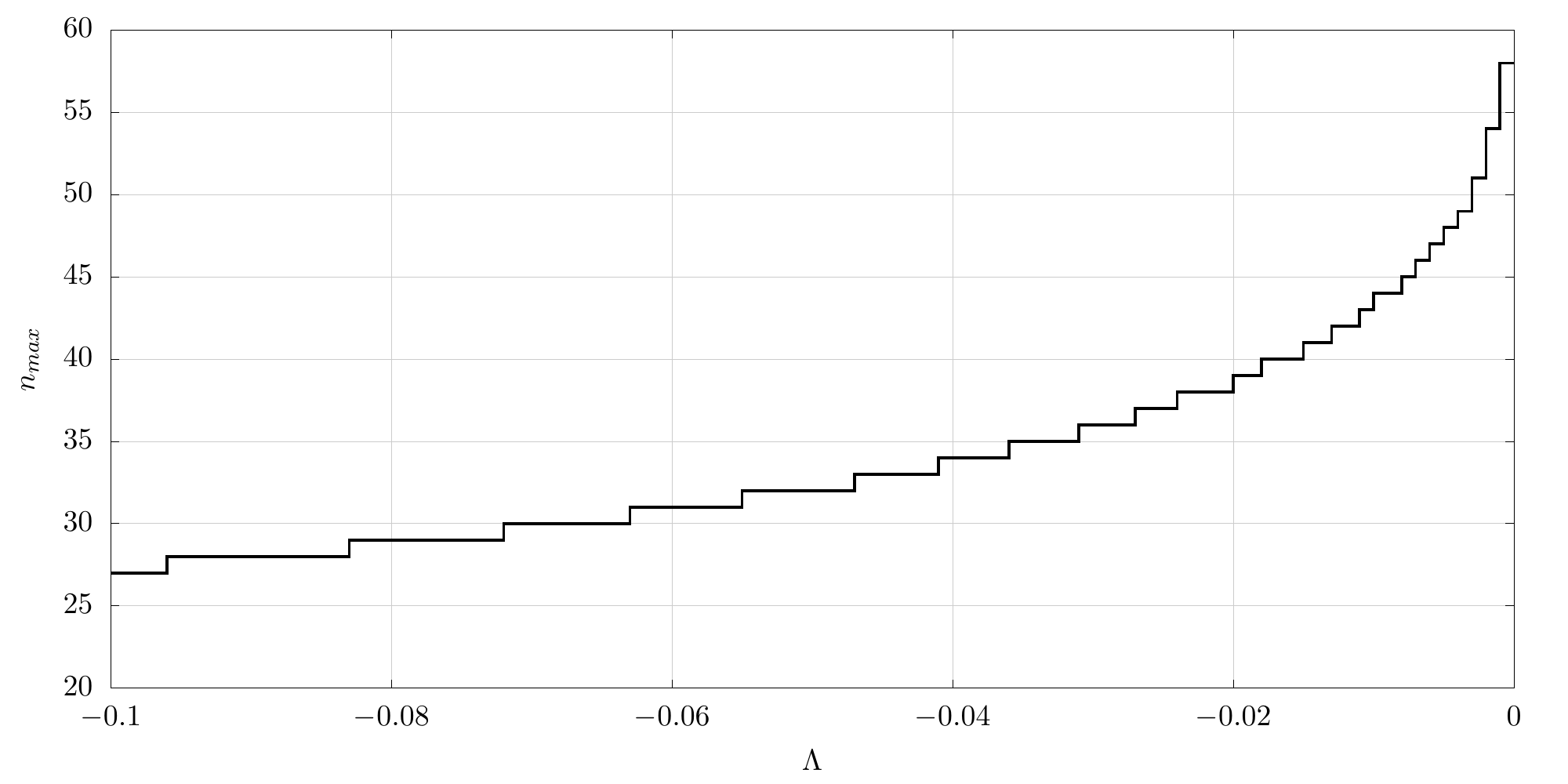}
\caption{The maximal value of topological charge soliton for the old baby potential with $\lambda=\mu=1$ and $\kappa^2=0.1$, for different $\Lambda$. }
\label{maxN}
\end{figure}

Although the higher charge axially symmetric solutions probably are not the true energy minimisers, we use them to get some intuition about the impact of the cosmological constant on the ADM mass-radius curve. Concretely, we plot this curve for the old baby potential which, as we know, supports compactons. Hence, the geometric volume (radius) is a well defined quantity. We chose parameters $\lambda=\mu=1$ and $\kappa^2=0.1$ such that in the $\Lambda=0$ limit $n<n_{\max}=60$.  
We find that $R_{max}$ and $M_{max}$ increase for increasing $|\Lambda|$, see Fig. \ref{MR-AdS-1}.
Qualitatively, the role of the cosmological constant is clear. It acts as negative pressure i.e., an additional repulsive interaction which counterbalances the gravitational attraction. Therefore, for a given value of  the topological charge, we find less compressed (gravitationally squeezed) matter with bigger radius. Furthermore, solutions exist for higher masses as the negative $\Lambda$ has an opposite effect on the appearance of the singularity of the metric than the matter density. Qualitatively, with increasing $|\Lambda|$ the mass-radius curve has a tendency to approach the analytical relation found for $\kappa=0$ (\ref{MR-kappa0}) - dashed line in the plot. Of course, in this limit there is no upper limit on the mass (radius) and the curve continues to infinity. 
Although the maximal ADM mass grow with $|\Lambda|$, the maximal topological charge of the baby Skyrmion decreases, as is shown in Fig. \ref{maxN}. 


\begin{figure}
%\hspace*{-1.0cm}
\includegraphics[height=5.5cm]{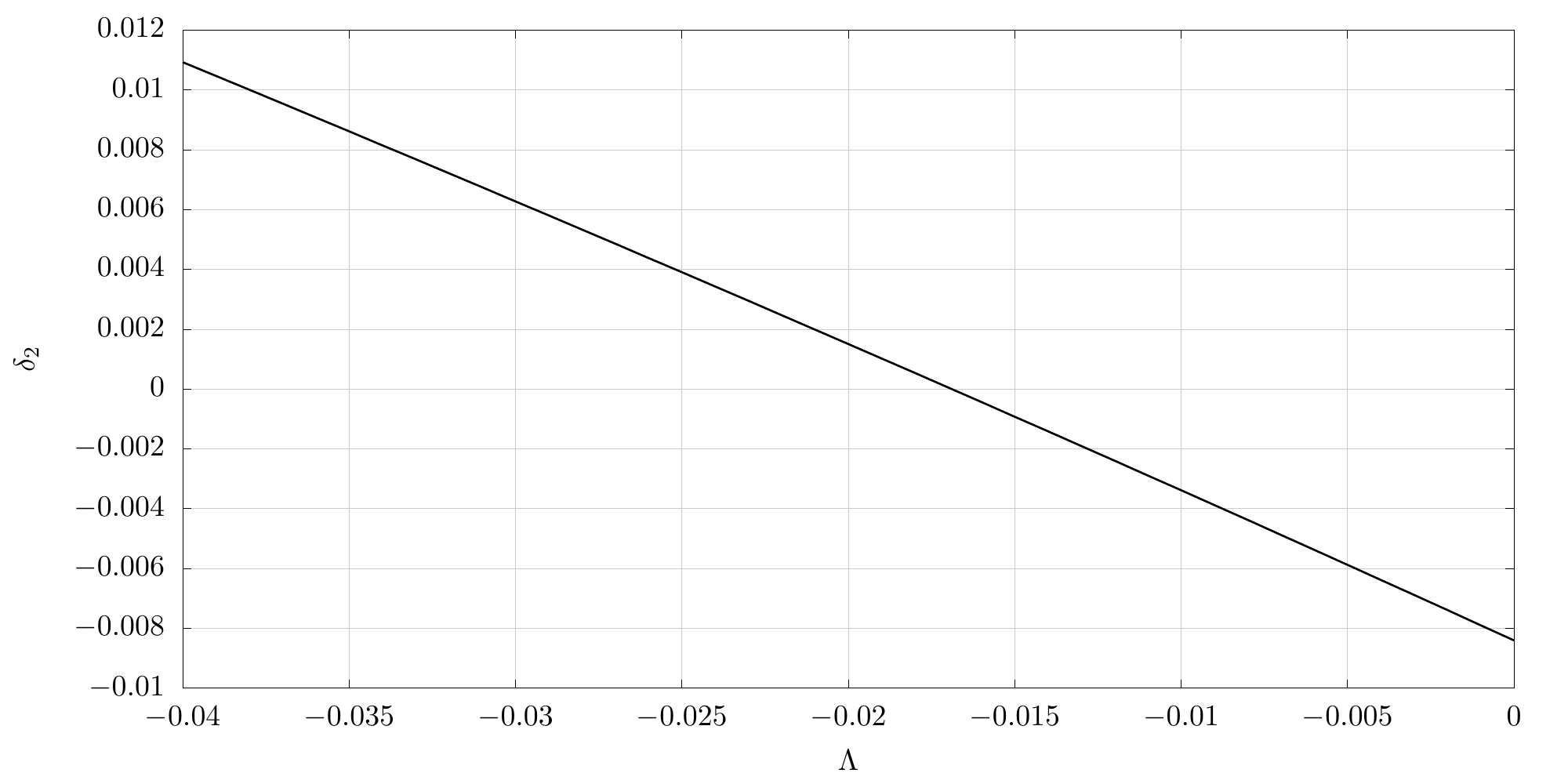}
\caption{The relative excess energy for the old baby potential for different values of $\Lambda$. Here $\lambda=\mu=1$ and $\kappa^2=0.1$. }
\label{E-AdS}
\end{figure}

\begin{figure}
%\hspace*{-1.0cm}
\includegraphics[height=6.5cm]{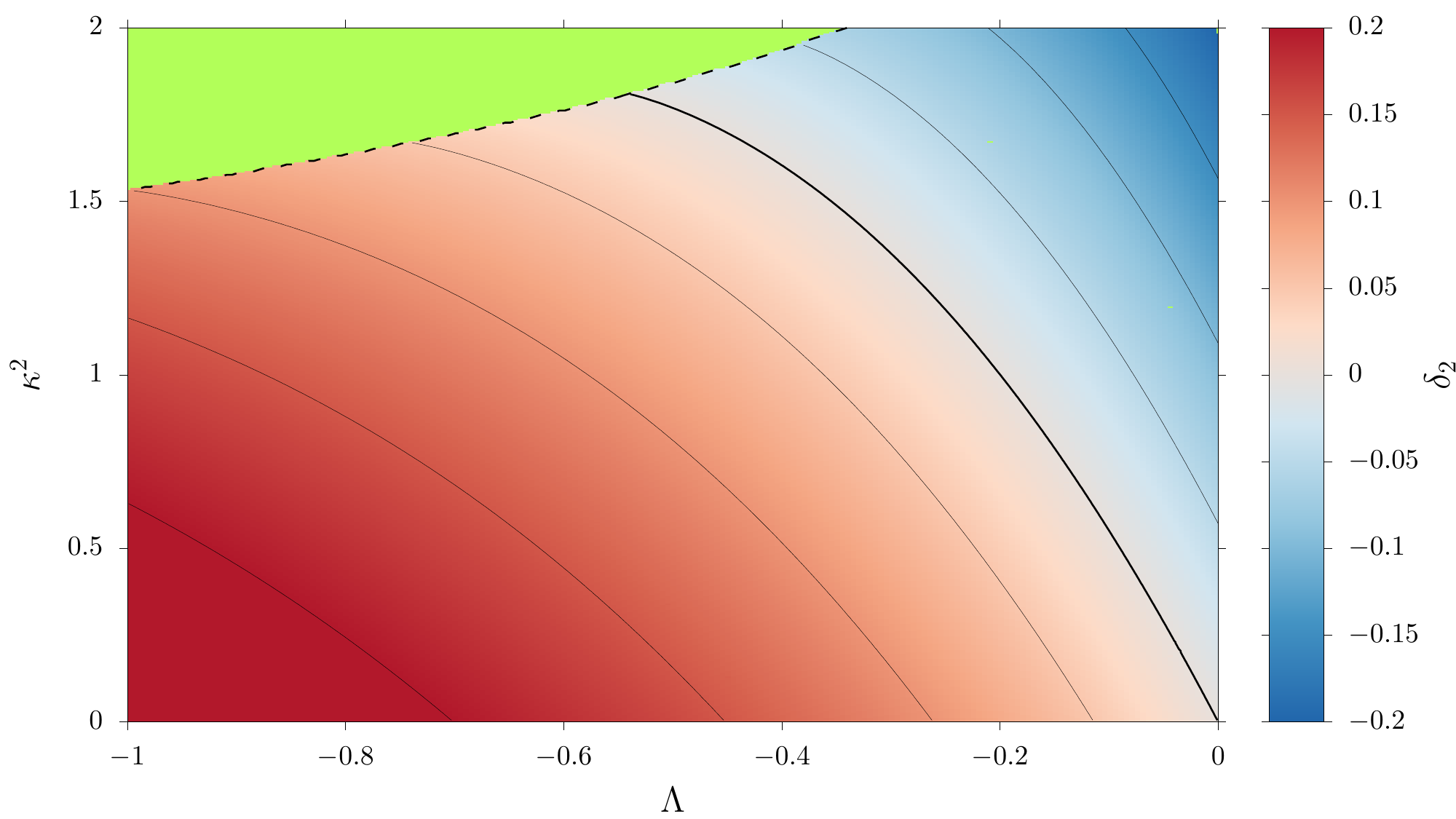}
\caption{Regions of positive and negative excess energy $\delta_2$ in the $\Lambda$-$\kappa^2$ plane. The green region close to the upper left corner corresponds to parameter values where a $n=2$ soliton solution does not exist, i.e., $ n_{\rm max} =1$. Here $\lambda=\mu=1$.}
\label{delta2scan}
\end{figure}

Next, for the old baby potential with the same choice of the parameters $\mu=\lambda=1$ and $\kappa^2=0.1$ we plot the relative excess energy 
\be
\delta_2=\frac{M_{ADM}(n=2)}{2M_{ADM}(n=1)} -1
\ee 
as a function of the cosmological constant - see Fig. \ref{E-AdS}.   It should be underlined that this is not just minus the relative binding energy. This quantity does not take into account the fact that a translation of a compacton in AdS space changes its energy. Hence, a set of two charge one compactons, sufficiently separated to form a non-overlapping pair, do not have an energy equal to twice the charge one energy. Therefore, a positive energy excess does not necessary imply the unstability of a self-gravitating multi-soliton solution. On the other hand, if this quantity is negative we know that the charge two solution is stable. As we remember, in the asymptotically flat space-time ($\Lambda=0$) this quantity is always (for any $\kappa$) negative, while in the no-backreaction limit ($\kappa=0$) in the AdS space-time it is always positive (for any $|\Lambda|$). This suggests that there can be a transition from  negative to positive (relative) excess energy while $\kappa$ and $\Lambda$ are changed. This is exactly what we find in Fig. \ref{E-AdS}. In particular, the critical value for the cosmological constant for which the binding energy changes its sign reads $\Lambda=-0.017$. A general behavior is that bigger $|\Lambda|$ moves $\delta_2$ towards bigger (positive) values. Hence, if we start with $\Lambda=0$ and a chosen value of the gravity coupling $\kappa$, the excess energy is negative. Then, if we increase $|\Lambda|$, it linearly grows to $\delta_2=0$. Above this value of the modulus of the cosmological constant the excess energy is always positive. 
Finally, in Fig. \ref{delta2scan} we scan for positive and negative excess energies $\delta_2$ in the $\Lambda$-$\kappa^2$ plane.

\begin{figure}
%\hspace*{-1.0cm}
\includegraphics[height=5.5cm]{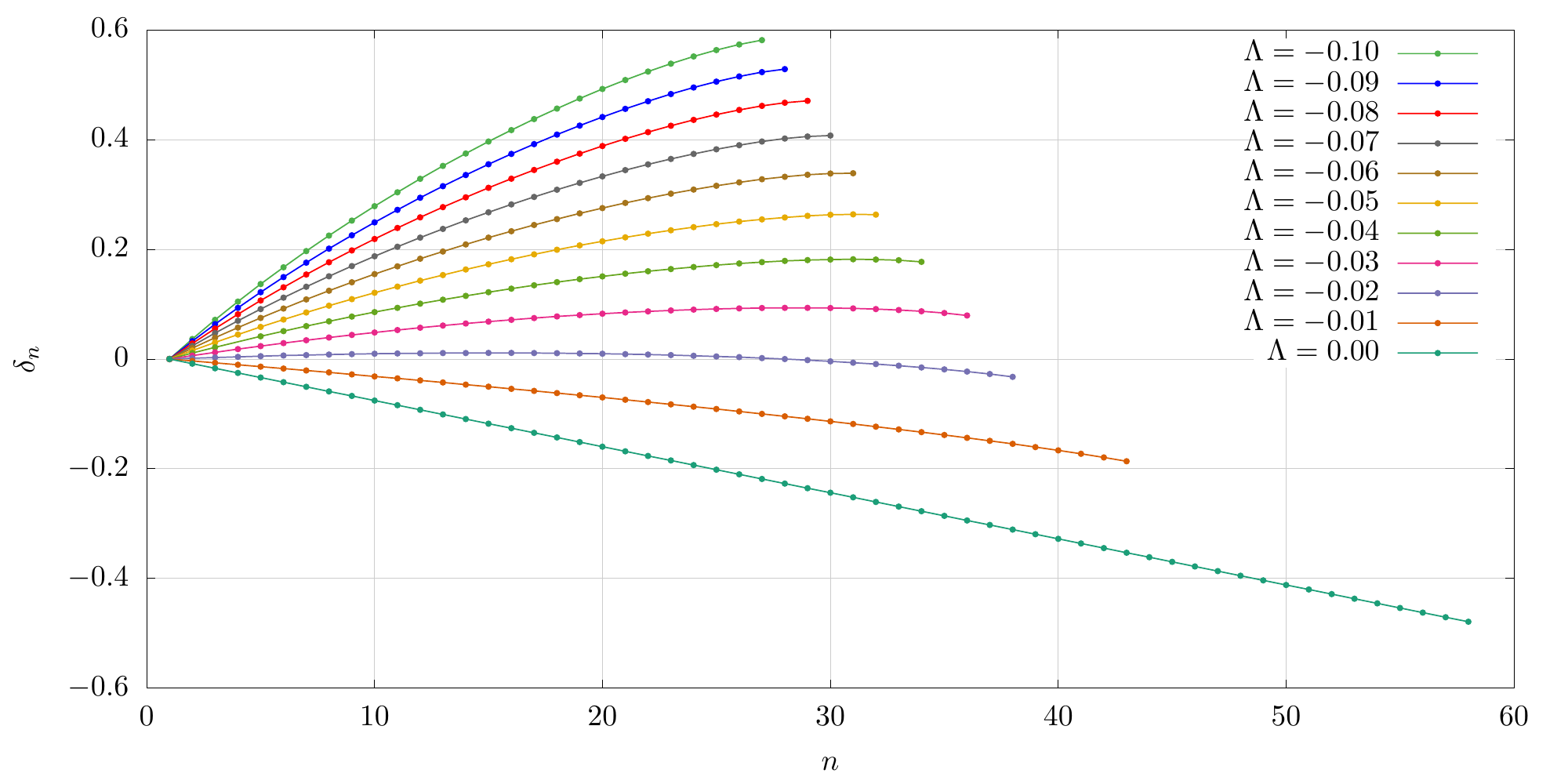}
\caption{The relative excess energies for the allowed values of the topological charge for the old baby potential for different values of $\Lambda$. Here $\lambda=\mu=1$ and $\kappa^2=0.1$. }
\label{EB-n-AdS}
\end{figure}

This pattern can be studied in a more elaborated way if we compute the relative excess energy for any allowed topological charge
\be
\delta_n=\frac{M_{ADM}(n)}{nM_{ADM}(1)} -1
\ee
In Fig. \ref{EB-n-AdS} we show our numerical findings. For $\Lambda=0$ all solitons have negative excess energy and higher charge Skyrmions are more tightly bound than the lighter ones. This is clearly explained by our results for the asymptotically flat space-time. In this case, the excess energy of a charge $n$ solution is simply a linear function of the topological charge
\be
\delta_n=-\frac{1}{2}  \frac{ \kappa^2 \lambda \mu   \langle \sqrt{\mathcal{U}} \rangle_{\mathbb{S}^2}}{1- \frac{1}{4} \kappa^2 \mu \lambda  \langle \sqrt{\mathcal{U}} \rangle_{\mathbb{S}^2}} |n| .
\ee 
Once we add a negative $\Lambda$ then the excess energy is less and less negative. However, as in the $\Lambda=0$ case the excess energy is proportional to the topological charge, the lightest solitons are affected first. Indeed, for small negative $\Lambda$ we observe that the first few solitons have {\it positive} excess energy while heavier baby Skyrmions still possess negative excess energy. The number of multi-solitons with negative excess energy decreases with growing $|\Lambda|$. For sufficiently large $|\Lambda | $ all solitons have positive excess energy. 

\begin{figure}
%\hspace*{-1.0cm}
\includegraphics[height=6.5cm]{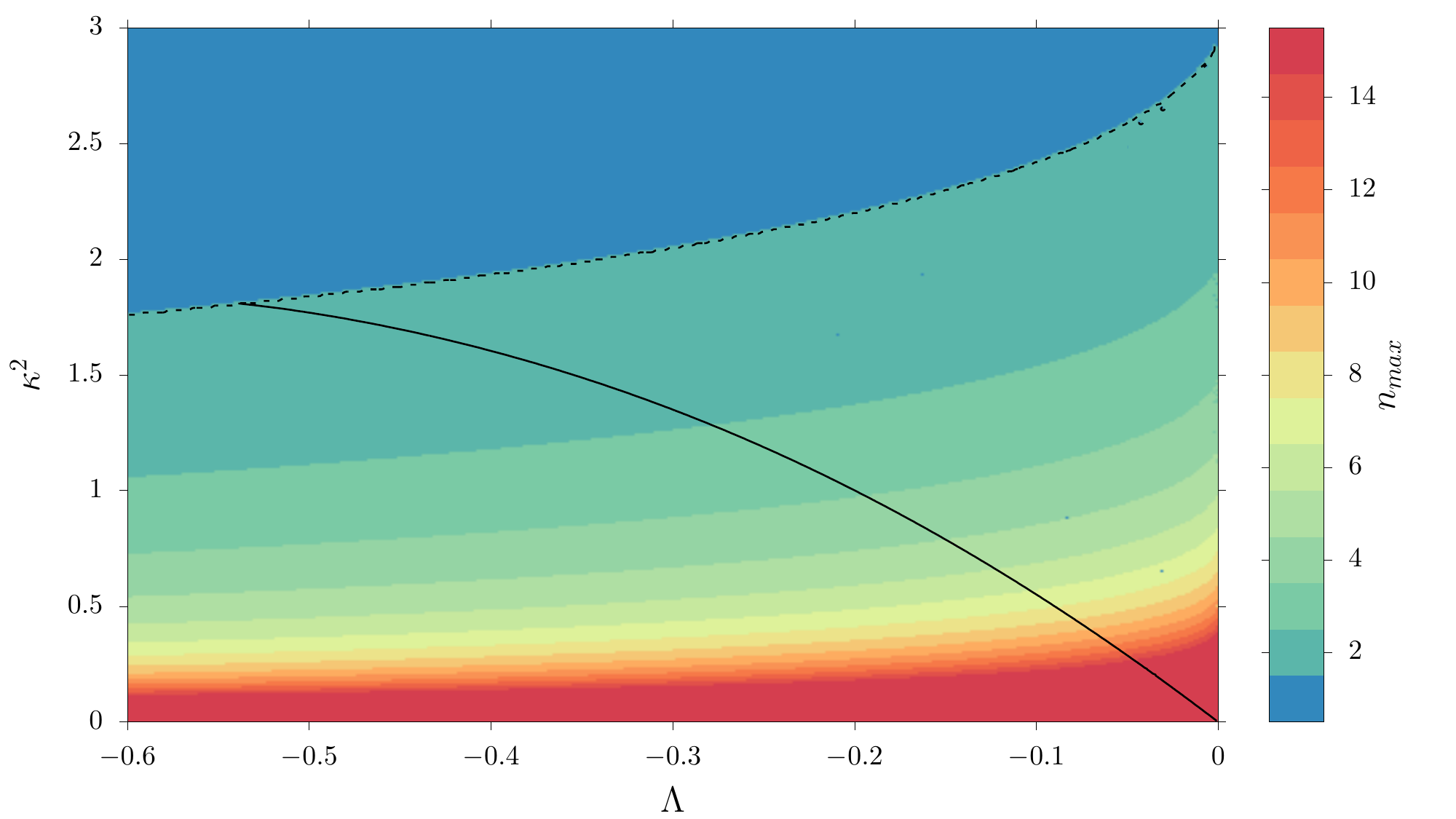}
\caption{ $n_{\rm max}$ as a function of $\Lambda$ and $\kappa^2$, for $1\le n\le 15$. In the blue region, $n_{\rm max} =1$. The continuous black line separates positive and negative energy excess  for charge $n$ solitons. Here $\lambda=\mu=1$. }
\label{nmax-fig}
\end{figure}

Finally, we plot $n_{\rm max}$ as a function of $\Lambda$ and $\kappa^2$ in Fig. \ref{nmax-fig}.

%%%%%%%%%%%%%%%%%%%%%%%%%%%%%%%%%%%%%%%%%
\section{Summary}
%%%%%%%%%%%%%%%%%%%%%%%%%%%%%%%%%%%%%%%%%
In the present paper, we investigated the baby BPS Skyrme model after coupling with gravity in $(2+1)$ dimensions. From the point of view of the energy-momentum tensor, this skyrmionic matter system still describes a perfect fluid, where each potential provides a particular (non-barotropic) equation of state. 

For the asymptotically flat metric, we found that the model still exhibits the BPS property. As a result, there are solitonic solutions obeying a Bogomolnyi equation, i.e., a local zero-pressure equation, for arbitrary values of the winding number. After a change of the radial coordinate, this Bogomolnyi equation exactly coincides with the Bogomolnyi equation of the BPS baby Skyrme model without gravity. As a result, the proper (non-gravitational) mass of the solutions is a linear function of the topological charge, with the coefficient determined by a target space integral. Analogously, the proper geometrical volume (for compactons) is a linear function of the topological charge. A generic impact of gravity is visible in the appearance of a maximal mass, corresponding to a maximal topological charge, allowed for gravitating BPS baby Skyrmions. Furthermore, we found exact formulas for the total mass and radius. Both get negative contributions due to the gravitational interaction which are quadratic in the topological charge. We observed that the mass instability happens exactly at the maximal mass point. Finally, we obtained an exact total (asymptotic) mass-radius curve which qualitatively agrees with its higher dimensional counterpart. We want to underline that all results are obtained in a completely analytical manner, which implies the exact solvability of this  gravitating system. Note that, although it is perfectly possible to compute the local form of solutions, all quantities are expressed via target space integrals of functions of the potential. Especially, the shape of the mass-radius curve is analytically related to a ratio between target space integrals of the potential (with compact solutions). Therefore we  completely characterise the mass-radius curve (or more generally properties of "toy neutron stars") by properties of the underlying matter theory, that is, a particular form of the potential ("toy nuclear matter"). Quite interestingly, there can be (infinitely) many distinct potentials which give exactly the same mass-radius curve. Such a degeneracy puts some doubts on the so-called inverse problem, i.e., the reconstruction of the equation of state (which corresponds to a potential in the language of the effective Skyrme theory) from the $M$-$R$ curve. 

\vspace*{0.2cm} 

Some of these findings resemble the situation of the gravitating $O(3)$ $\sigma$-model in the asymptotically flat space-time (although here we have infinitely many models defined by different potentials). Therefore, the full gravitating baby Skyrme model can still be understood as a sum of two separate BPS parts - each with its own Bogomolnyi equation and topological bound. 

Let us remark that the preservation of the BPS property in the model after its coupling to gravity most likely implies the existence of an $\mathcal{N}=2$ supersymmetric extension of the BPS baby Skyrme model \cite{susy} also with gravity included. 

\vspace*{0.2cm} 

The model gets more complicated once a non-zero cosmological constant is added. In the case of positive $\Lambda$ (de Sitter space-time) one can distinguish two different, {\it extremal} and {\it non-extremal},  types of solutions. The extremal solution is a local constant pressure solution and, thus, closely related to the non-gravitating model. In fact, after a suitable change of the radial coordinate one can cast the matter equation in the form of the Bogomolnyi equation in flat space-time, generalised for constant non-zero external pressure, where now the role of the external pressure is played by the cosmological constant. Again, the mass and volume of the baby Skyrmions  can be found in an exact way. Such a solution is well-defined only if it covers the whole allowed space. Equivalently, one can say that the metric must possess a singularity at the boundary of the extremal baby Skyrmion. This leads to a  relation between the parameters of the model and the topological charge. This relation was obtained in a closed form for any model (any potential).  

When the parameters of the model do not obey this relation, we can still have non-extremal solutions. Now, the pressure changes inside the solitons, and the solutions lose any similarity with the non-gravity case. 

\vspace*{0.2cm} 

There are no extremal solutions for the asymptotically anti-de Sitter space-time. Such solutions are forbidden, as there are no constant negative pressure solutions in the non-gravity case with non-zero topological charge. Again, gravitating solitons lose any similarity with the original flat space-time BPS baby Skyrmions. 

Interestingly enough, there are two analytically solvable cases where analytical axially symmetric multi-solitons in the external AdS space-time can be found. It includes the model with the old baby potential, leading to compact solutions.  
The axially symmetric higher charge solutions correspond to a liquid phase where individual charge one solitons get completely dissolved. In (asymptotically) flat space, there exists a second, gaseous phase of non-overlapping charge one solitons. It is an interesting question whether such a second phase continues to exist in AdS space, and, in the affirmative case, which phase has lower energy.   
Here we made a first step towards answering this question by computing the excess energy, which is the difference between the mass of the charge two axially symmetric solution and twice the energy of the charge one axially symmetric solution. Note that this does not take into account the additional energy coming from the fact that a state of  two separated compactons in AdS space-time requires to move the solitons away from the origin. This excess energy is positive for external AdS that is when $\kappa=0$ and decreases to negative values while $\kappa$ grows. Obviously, if it becomes negative then the fluid phase is the true energy minimum configuration. On the other hand, to study the implications of a positive excess energy for the existence and stability of different phases requires full two-dimensional numerical calculations.
However, even if higher charge solitons are not completely destabilised by positive excess energies, these positive excess energies still may reduce total binding energies and, thereby, provide an explanation of why holographic baby Skyrmions in an \textit{AdS} background in the full baby Skyrme model have small binding energies \cite{sut-ads}.

\vspace*{0.2cm}

Looking from a wider perspective, we can conclude that there is an intimate relation between the non-gravitating BPS baby Skyrmions in flat space-time without (with) pressure and properties of their gravitating counterparts in an asymptotically flat (curved) space-time. 

%%%%%%%%%%%%%%%%%%%%%%%%%%%%%%%%%%%%%%%%%
\section*{Acknowledgements}
%%%%%%%%%%%%%%%%%%%%%%%%%%%%%%%%%%%%%%%%%
The authors acknowledge financial support from the Ministry of Education, Culture, and Sports, Spain (Grant No. FPA 2014-58-293-C2-1-P), the Xunta de Galicia (Grant No. INCITE09.296.035PR and Conselleria de Educacion), the Spanish Consolider-Ingenio 2010 Programme CPAN (CSD2007-00042), Maria de Maetzu Unit of Excellence MDM-2016-0692, and FEDER. A.W. thanks Lukasz Bratek and Paul Sutcliffe for comments.

\end{document}